\documentclass[11pt]{article}

\usepackage{amssymb}
\usepackage{amsfonts}
\usepackage{amsmath}
\usepackage{color}
\usepackage[pdftex]{graphicx}
\usepackage{lscape}
\usepackage{bm}
\usepackage{url}
\usepackage{textgreek}

	\newtheorem{theorem}{Theorem}
	\newtheorem{assumption}{Assumption}
	
	\newtheorem{corollary}{Corollary}
	
	\newtheorem{lemma}{Lemma}
	\newtheorem{proposition}{Proposition}
	
    \newtheorem{definition}{Definition}
    \newtheorem{property}{Property}

\begin{document}

\title{\textsc{Predicting the Unpredictable under Subjective Expected Utility}\thanks{I thank J\"{u}rgen Eichberger, Ani Guerdjikova, Stephan Jagau, Rabee Tourky, Peter Wakker, and participants at the Mini-Workshop on Decision Theory with Unawareness and the annual conference of the Verein f\"{u}r Socialpolitik 2022 for very helpful discussions. This paper grew out of a current joint project with Stephan Jagau on experimentally testing subjective prediction rules for novelties. Financial support via ARO Contract W911NF2210282 is gratefully acknowledged.}}

\author{Burkhard C. Schipper\thanks{University of California, Davis. Email: bcschipper@ucdavis.edu}}

\date{First Version: February 26, 2022. This Version: February 18, 2025.}

\maketitle

\begin{abstract} We consider a decision maker who is unaware of objects to be sampled and thus cannot form beliefs about the occurrence of particular objects. Ex ante she can form beliefs about the occurrence of novelty and the frequencies of yet to be known objects. Conditional on any sampled objects, she can also form beliefs about the next object being novel or being one of the previously sampled objects. We characterize behaviorally such beliefs under subjective expected utility. In doing so, we relate ``reverse'' Bayesianism, a central property in the literature on decision making under growing awareness, with exchangeable random partitions, the central property in the literature on the discovery of species problem and mutations in statistics, combinatorial probability theory, and population genetics. Partition exchangeable beliefs do not necessarily satisfy ``reverse'' Bayesianism. Yet, the most prominent models of exchangeable random partitions, the model by De Morgan (1838), the one parameter model of Ewens (1972), and the two parameter model of Pitman (1995) and Zabell (1997), do satisfy ``reverse'' Bayesianism. Our characterization allows us to interpret these models as subjective beliefs of a decision maker and to derive the parameters from choice behavior.\\
\\
\noindent {\bf Keywords}: Awareness of unawareness, unknown unknowns, exchangeable random partitions, ``reverse'' Bayesianism, discovery of species problem, discovery, novelty, inductive reasoning.\\

\noindent {\bf JEL Classification Numbers}: D83
\end{abstract}

\newpage

\section{Introduction}

Consider a researcher/funding agency/humanity who each period creates/funds/invents a(n) idea/research project/technology. How should such an agent form beliefs about whether or not the outcome is truly novel or a derivative of an existing idea/project/technology? The issue is that ex ante neither the researcher, funding agency, or humanity is aware of all research outcomes or technologies. Therefore it is impossible to assign subjective probabilities to particular future ideas, research outcomes, or technologies. Yet, the decision maker can surely reason about events like the ``occurrence of a novelty in period $t$'', the ``reoccurrence of the outcome in period $t+k$ that was previously encountered in period $t$'' etc. Such reasoning involves propositions about objects that are stripped of their concepts. Their names are ``delabeled'' and objects are labeled instead by their first time of occurrence. These arguments suggest that the decision maker is at least able to reason about sampling times and possible partitions of sampling times into equivalence classes in which the same yet to be known outcome is encountered.

\begin{figure}[h!]
\caption{Partition of Sampling Times till $T = 4$\label{T4}}
\begin{center}
\includegraphics[scale=0.4]{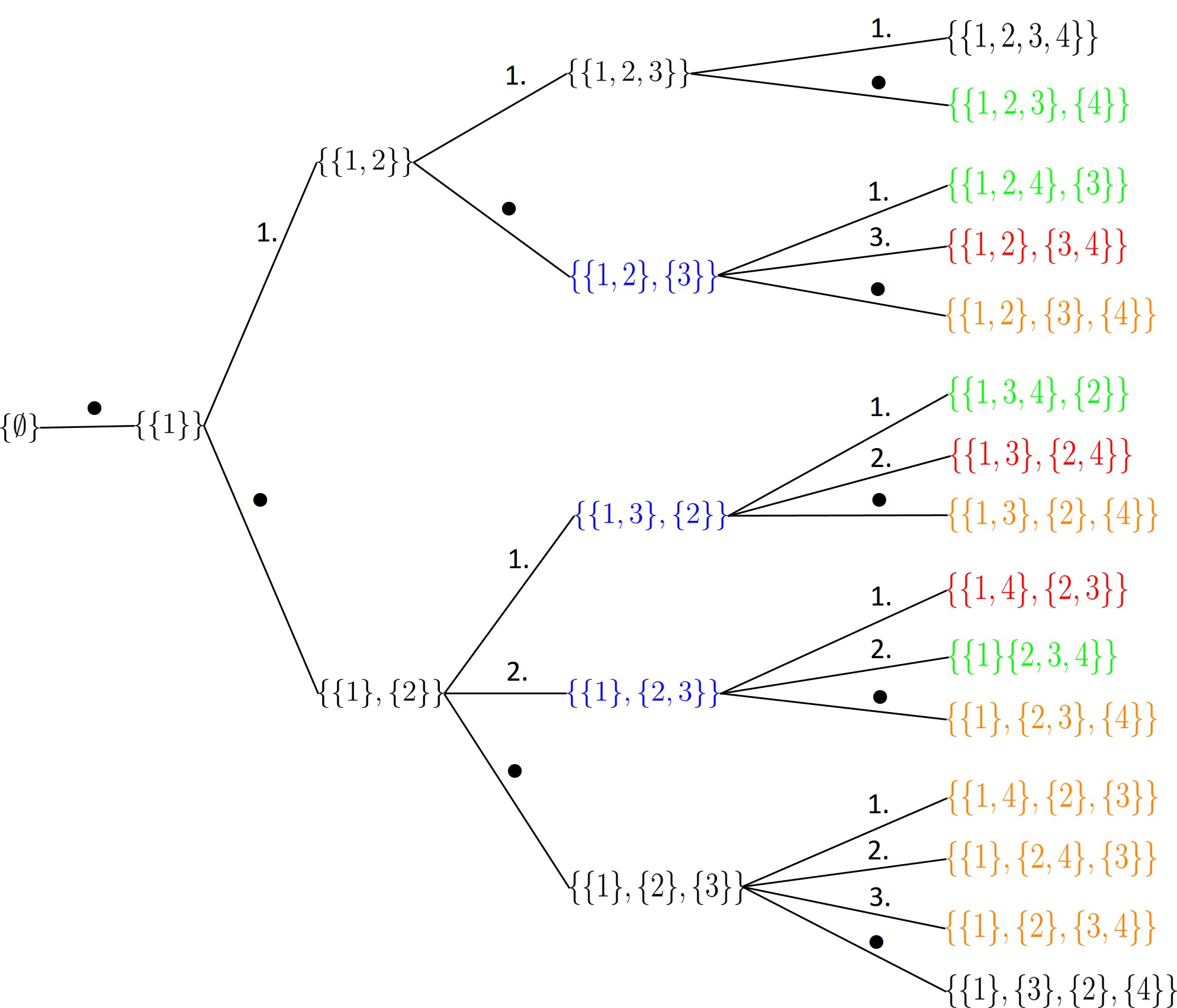}
\end{center}
\end{figure}

To fix ideas about the setting, consider Figure~\ref{T4}. It depicts the evolution of partitions of sampling times till period $T = 4$. Initially, nothing has been sampled as indicated by the empty set $\emptyset$ at the left side of the figure. Since the decision maker is unaware of any particular objects, she can only consider that a novel object will be drawn in period 1. We indicate the event ``a novel object is drawn'' by a black ball ``$\bullet$'' above the edge. This leads to the trivial partition of sampling times in period $T = 1$, namely $\{\{1\}\}$. Now the agent is aware of the first object. She can consider the event that the first object is drawn again, indicated by ``1.'' on top of the upper edge connecting partition $\{\{1\}\}$ to partition $\{\{1, 2\}\}$, or that a (different) novel object is drawn, indicated by ``$\bullet$'' on top of the lower edge connecting $\{\{1\}\}$ to partition $\{\{1\}, \{2\}\}$. The possible partitions at period $T = 2$ are $\{\{1, 2\}\}$, i.e., a first object was drawn in the first period and it was redrawn in the second period, and $\{\{1\}, \{2\}\}$, i.e., a first object was drawn in period 1 and a second distinct necessarily novel object was drawn in period 2. The process continues. Whenever the second object is redrawn it is indicated by ``2.'' above the edge. Analogously we indicate the third object with ``3.''.

What structure should be imposed on such beliefs and their evolution? We are not the first to consider such questions. Related questions have been considered in the literature on the discovery of species problem and population genetics, and the subsequent literature in statistics, combinatorial probability theory, and mathematical biology. Following Kingman (1978a, b), Aldous (1985), Pitman (1995, 2006), Zabell (1992, 1997) and others, we may impose partition exchangeability, an extension of de Finetti's exchangeability idea to partitions. Intuitively, suppose that each time sampling is performed in a similar fashion. That is, each period the process of organizing/funding a different avenue of research follows the same procedure. Then it might be justified to surmise that the occurrence of a known object in one period followed by a novelty in the next period is as likely as the occurrence of a novelty in the first period and a known object in the next period. More generally, partitions of sampling times that have the same frequency of partitions cells of the same cardinality (e.g., frequencies of frequencies) should be judged equally likely. In terms of Figure~\ref{T4}, it means that partitions printed in the same color (apart from black) should have the same probability. For instance, all the blue partitions have exactly a cell of cardinality one and a cell of cardinality two and no other partitions have exactly these features. In our decision theoretic framework, we impose a symmetry property on preferences over Savage acts that will characterize partition exchangeable subjective beliefs.

We are interested in updating of beliefs conditional on novel objects being drawn. What are the predictive subjective probabilities for drawing another novel object or for redrawing a particular object drawn earlier? For instance, we may want to impose that the likelihood of novelty should only depend on how often novelties have occurred in the past and on how many periods we have sampled so far. Moreover, we may want to impose that the likelihood of a known object should only depend on its frequency so far and on how many periods we have been sampling. In such a case, Zabell (1997) showed that the prediction rule is characterized by two parameters. Consider sampling period $T$. The conditional probability of a novelty is $$\mbox{Cond. prob. of a novel object in } T+1 = \frac{\alpha \cdot \mbox{Number of times novelties experienced } + \theta}{T + \theta}$$ and the conditional probability of the $j$th distinct object that occurred previously is $$\mbox{Cond. prob. of the }j\mbox{th distinct known object in } T+1 = \frac{\mbox{Number of times object }j\mbox{ occurred} - \alpha}{T + \theta}$$ with $\alpha \in [0, 1)$ and $\theta > - \alpha$. We characterize this prediction rule in terms properties on preferences over Savage acts. If subjective beliefs conform to the prediction rule, then behavior must correspond to the behavioral characterization and vice versa. From a positive point of view, it allows for testing with choice experiments whether or not a decision maker entertains such subjective beliefs. From a normative point of view, it suggests guidelines on behavior that need to be satisfied if and only if the prediction rule is judged desirable. The behavioral characterizations of the prediction rules allow us to endogenously derive the parameters, which are exogenously given in the theory of random partitions, from choice behavior. The parameter $\theta$ is positively associated with the prior belief of a novel object. The parameter $\alpha$ is positively associated with the effect of observing novelty in the past on the belief of future novelty. We can think of the decision maker as keeping a ``mental'' urn consisting of colored balls and a black ball. Colored balls stand for known objects and the black ball for novelty. Initially the urn contains a black ball with weight $\theta$. Each time a colored ball is drawn, it is returned to the urn together with another ball of the same color. Each time a black ball is drawn, it is returned to the urn together with two other balls, one black ball with weight $\alpha$ and one of a new color with weight $1 - \alpha$. See Aldous (1985) and Pitman (2006, Sections 3.1-2) for alternative interpretations with the Chinese restaurant process.

The two-parameter model has been central to the literature on exchangeable random partitions (see Pitman, 1995). It generalizes the ubiquitous one-parameter model due to Ewens (1972) according to which the conditional probability of a novelty only depends on the sampling time. In this case, $\alpha = 0$, and the urn interpretation corresponds to the Hoppe urn (Hoppe, 1984); see Crane (2016) for a survey on the model associated with the Ewens sampling formula. We also provide properties on preferences that characterize the Ewens model. In the special case, when $\theta = 1$, this is the De Morgan (1838, 1845) prediction rule, which is likely the oldest explicit prediction rule for novelty; see Zabell (1992, 1997) for marvelous discussions.

Well-justified prediction rules for novelties are of utmost importance for humanity. Novelties can be both the cause for demise and the source for opportunities for humanity. Given the short written human history, we just have limited data on human creativity. Thus, a subjective interpretation of novelty creating processes seem desirable. For instance, the research movement on existential risk aims to predict the occurrence of novel events that could devastate civilization and ways to mitigate such risks. Typically researchers state subjective risk assessments and seem to be open to arguments with ``mental'' urn models. For instance, Bostrom (2019) in his discussion of the vulnerable world hypothesis uses an urn analogy to grasp the process of human creativity. A behavioral characterization of subjective beliefs about future novelties allows for interpreting them as bets.

While our analysis of the evolution of subjective beliefs in the face of discovery of novel objects draws heavily on the statistical literature motivated by the discovery of species problem and population genetics, our motivation comes from the literature on subjective belief updating under unawareness. Unawareness refers to the lack of conception rather than the lack of information. A decision maker is unaware of an object if she cannot conceive of it. However, this does not preclude her from realizing that there might be objects out there that she cannot conceive of. This refers to awareness of unawareness. Several unawareness logics have been developed that provide a rich formal framework for modeling unawareness\footnote{See Fagin and Halpern (1988), Modica and Rustichini (1999), Heifetz, Meier, Schipper (2005, 2008, 2013), Halpern and R\^{e}go (2008), and Li (2009). See Schipper (2015) for a survey.} and even awareness of unawareness.\footnote{See Board and Chung (2021), Halpern and R\^{e}go (2009, 2013), Sillari (2008), {\AA}gotnes and Alechina (2014), Halpern and Piermont (2019), Fukuda (2021), and Schipper (2024).} Unawareness is not just a topic of intellectual curiosity but has been applied to game theory, economics, contract theory, finance, business strategy, electoral campaigning, strategic network formation etc.\footnote{For a bibliography, see \url{http://faculty.econ.ucdavis.edu/faculty/schipper/unaw.htm}.}

In terms of belief updating in the face of growing awareness, Karni and Vier{\o} (2013, 2017) and Hayashi (2012) characterize behaviorally ``reverse'' Bayesianism\footnote{See also Dominiak and Tserenjigmid (2018, 2022), Dominiak (2022), Karni, Valenzuela-Stookey, and Vier{\o} (2021), Vier{\o} (2021, 2023), and Chakravarty, Kelsey, and Teitelbaum (2022). Dietrich (2018) also considers proportional changes of probabilities upon becoming aware.} according to which the relatively likelihood ratio for events that the decision maker is already aware of remains unchanged upon becoming aware of a novel event.\footnote{This principle has been experimentally studied by Becker et al. (2022). It has been generalized by Piermont (2021). It has been applied as a belief restriction to solutions in games with unawareness by Francetich and Schipper (2023). De Canson (2024), Steele and Stef\'{a}nsson (2021), Mahtani (2021), Roussos (2021), Bradley (2017), and Wenmackers and Romeijn (2016) provide further discussions of the principle.} Karni and Vier{\o} (2013) and Hayashi (2012) did not consider awareness of unawareness, i.e., beliefs about becoming aware. However, in follow-up work, Karni and Vier{\o} (2017) extend their approach to beliefs about becoming aware (see also Vier{\o}, 2021, for an intertemporal setting, and Vier{\o}, 2023, for a setting with potentially different unknowns). They just consider the case of becoming aware of another outcome while we consider the case of becoming aware of states/objects. We show in Section~\ref{RB} that partition exchangeable beliefs do not necessarily satisfy ``reverse'' Bayesianism. Yet, the two-parameter model by Pitman (1995) and Zabell (1997), the Ewens' one-parameter model, and the De Morgan model do satisfy it. They can be used to come up with more precise ``reverse'' Bayesian updates in sampling problems. We also show that they satisfy an extended notion of Bayesianism involving the invariance of likelihood ratios for sampling novelty versus a known object upon sampling another known object. We also observe that the prediction rule for novelties by Kuipers (1973) (that is inconsistent with partition exchangeability) does not satisfy ``reverse'' Bayesianism.

The idea of bridging subjective expected utility theory and inductive methods when sampling objects is inspired by Wakker (2002). He characterizes subjective expected utility theory in the context of sampling known objects that satisfies Carnap's prediction rule (also called Carnap's continuum of inductive methods or Carnap's inductive logic). The two-parameter model characterized in our paper can be viewed as a generalization of Carnap's prediction rule allowing also for novel objects; see Zabell (1997) for a discussion. Both Wakker (2002) and we make use of Zabell's work on inductive methods. Wakker (2002) draws on Zabell's earlier characterization of the Carnap-Johnson prediction rule for known objects while we draw on Zabell's (1997) characterization for the prediction rule of the two-parameter model for sampling with novel objects. We also make use of Wakker's (1989) approach to subjective expected utility, in particular the extension of his ``tradeoff consistency'' approach to settings with countable additive probability measures. Two drawbacks of Savage's approach to subjective expected utility are that it just yields non-atomistic finitely additive probabilities measures. These features do not fit with our setting. We avoid them with Wakker's approach. Recently, Karni (2024) presented an extension of the Ewens model to sampling from several different urns that contain also novel outcomes and embeds it into a decision problem. In his approach, predictive probabilities are treated as objective probabilities while we characterize them as subjective probabilities. 

The paper adds to the decision theoretic literature on unawareness beyond Karni and Vier{\o} (2013, 2017) and Hayashi (2012).\footnote{Ahn and Ergin (2010) present an account of partition-dependent expected utility in which likelihood ratios are invariant to the partition of the state-space. They also study a notion of completely overlooked events that is different from the notion of propositional unawareness. Note though that they consider events that are unions of partitions cells while we consider partitions of sampling times as events, which is very different both at a formal and conceptual level.} Central to our approach is the idea of ``delabeling'' events that is behind the partition structures of sampling times in the statistical literature originating with Kingman (1978a,b). Even without concrete names or concepts, a decision maker can reason about the occurrence of distinct but yet to be labeled events over time. For that reason, we nicknamed Section~\ref{DSEU} ``delabeled subjective expected utility''. Our approach can be related to awareness-dependent subjective expected utility of Schipper (2013). He uses it to behaviorally characterize unawareness of events. The decision maker is unaware of an event if the event is null and the complement of the event is null. Awareness-dependent subjective expected utility is best understood as a static approach while the approach presented in this paper has a dynamic flavor. In Section~\ref{ASEU}, we demonstrate how the present approach can give rise to subjective predictive probabilities in unawareness structures when state spaces in unawareness structures are interpreted as arising recursively from a process of sampling novel objects. Schipper (2013) only considers unawareness proper but not awareness of unawareness. Kochov (2018) provides an approach to behaviorally reveal awareness of unawareness in a dynamic three-period model. Piermont (2017) uses menu choices to reveal awareness of unawareness. Grant and Quiggin (2013) argue compellingly that decision makers should inductively infer from having become aware of events in the past that there might be still events out there of which they are unaware. Dietrich (2018) studies subjective expected utility \`{a} la Savage with unawareness. Schipper (2014) connects the choice-theoretic approach to the epistemic approach to knowledge and unawareness. Grant, Meneghel, and Tourky (2021) study learning upon becoming aware where newly discovered events are initially ambiguous and become less so with learning. They assume that the underlying stochastic process is a Dirichlet process. The models considered in our paper are closely related to a generalization, the Poisson-Dirichlet process (see Crane, 2016, for a survey). Eichberger and Guerdjikova (2023) consider case-based decisions when the decision maker is aware of her unawareness of some characteristics of the decision context. They characterize preferences including a degree of unawareness which is the decision weight placed on ``other characteristics''. Finally, Schipper (2021) introduces a self-confirming equilibrium concept for games in extensive form with unawareness that can be understood as a result of a learning and discovery process.

The next section introduces ``delabeled'' subjective expected utility. Section~\ref{SPE} contains the main results on the behavioral characterizations of the prediction rules. Section~\ref{RB} relates the prediction rules to ``reverse'' Bayesianism. An application to unawareness structures is briefly discussed in Section~\ref{ASEU}. Section~\ref{discussion} concludes with a discussions. Proofs are relegated to an appendix.

\section{``Delabeled'' Subjective Expected Utility\label{DSEU}}

Time periods are discrete and indexed by $t \in \{1, 2, ....\}$. Each period, the decision maker samples an object. Initially, the decision maker is unaware of all objects but is aware that there are objects. Once an object is sampled, she becomes aware of the object. Thus, at any point of time she is aware of all objects sampled at prior periods and the fact that there could be novel objects never sampled before. Although she is unable to ex ante form beliefs about particular objects or sequences of particular objects sampled, she can form beliefs over partitions of sampling in times, for which each partition cell collects the sampling times in which the same object is sampled. Thus, as in the species sampling literature influenced by Kingman (1978a, b, c), we take partitions of sampling times as primitives of our analysis and study the decision makers beliefs over those partitions.

For any $T = 1, ...$, let $\Pi^T$ be the set of all partitions of $\{1, ..., T\}$. A generic partition of $\{1, ..., T\}$ is denoted by $\bm{\pi}^T = (\bm{\pi}^T_j)_{j = 1}^k$ for some strict positive integer $k \leq T$, where $\bm{\pi}^T_j$ is the $j$'th partition cell in the order of appearance. Since the order of appearance of partition cells of samplings times will play role in our exposition, we write the collection of partitions as a sequence using ``$( ...)$'' rather than as a set ``$\{...\}$''. Denote by $\bm{\pi}^T(t)$ the cell of partition $\bm{\pi}^T$ that contains $t$. For $T' < T$, we write
$\bm{\pi}^T_{\mid T'} := \left\{\bm{\pi}^T(t) \cap \{1, ..., T'\}\right\}_{t \in \{1, ..., T'\}}$. In words, $\bm{\pi}^T_{\mid T'}$ is the partition $\bm{\pi}^T$ restricted to sampling times $\{1, ..., T'\}$ simply by ``cutting off'' all sampling times after $T'$. We say that a collection of partitions $(\bm{\pi}^T)_{T = 1, ...}$ is consistent if for any $T > T' \geq 1$, $\bm{\pi}^T_{\mid T'} = \bm{\pi}^{T'}$. An infinite partition $\bm{\pi}^{\infty}$ of $\mathbb{N}_+$ is generated by a consistent collection of partitions $(\bm{\pi}^T)_{T = 1, ...}$.

Let $\Omega$ be a nonempty topological state space endowed with a $\sigma$-algebra $\Sigma_{\Omega}$. In subjective expected utility (see for instance Appendix~\ref{Wakker}), the state space is just like an abstract probability space. The interpretation of events depends on the particular application of subjective expected utility. In our setting, we are interested in events that model partitions of sampling times. To this end, we introduce a collection of consistent surjective measurable ``partition functions'', $\{\pi^T: \Omega \longrightarrow \Pi^T\}_{T = 1, ...}$, such that for any $T = 1, ...$,
\begin{itemize}
\item[(0)] Measurability: $(\pi^T)^{-1}(\bm{\pi}^T) \in \Sigma_{\Omega}$ for every partition of sampling times $\bm{\pi}^T \in \Pi^T$.
\item[(i)] Consistency: For any $T' < T$, $\pi^{T'}(\omega) = \pi^T(\omega)_{\mid T'}$.
\end{itemize}
Condition (0) just says that every partition of sampling times gives rise to a measurable event via the functions $\pi^T, T \geq 1$. Condition (i) requires that the collection of functions $\pi^T, T \geq 1$, gives rise to consistent partitions of sampling times. That is, we exclude inconsistent sequences of partitions from the analysis as they are objectively impossible. Note that the state space needs to be large since the set of all partitions is uncountable. 

For any $T \geq 1$, $\bm{\pi}^T \in \Pi^T$, define the event $$[\bm{\pi}^T] := \{\omega \in \Omega: \pi^T(\omega) = \bm{\pi}^T\}.$$ Since $\pi^T$ is measurable for every $T \geq 1$, $[\bm{\pi}^T] \in \Sigma_{\Omega}$ for every $\bm{\pi}^T \in \Pi^T$ and $T$.

An real-valued (Savage) act is a measurable mapping $f: \Omega \longrightarrow \mathbb{R}$. An act is \emph{simple} if it is measurable w.r.t. some finite partition of $\Omega$. We abuse notation and denote by $x \in \mathbb{R}$ the constant act yielding $x$ for all $\omega \in \Omega$. Let $\mathcal{F}$ denote a set of acts including constant acts and simple acts.

Consider a preference relation $\succeq$ defined on $\mathcal{F}$. We denote by $\succ$ and $\sim$ the strict preference and indifference, respectively. The preference relation will be assumed to satisfy properties such as weak order (i.e., completeness and transitivity), some weak monotonicity property, some tradeoff consistency property, and continuity properties that capture countable additive subjective expected utility (Wakker, 1989). See Appendix~\ref{Wakker} for details.

For any event $E \in \Sigma_\Omega$ and acts $f, g \in \mathcal{F}$, we write $f_E g$ for the composite act defined by for all $\omega \in \Omega$, $$f_E g(\omega) := \left\{ \begin{array}{cl} f(\omega) & \mbox{ if } \omega \in E \\
g(\omega) & \mbox{ if } \omega \in \Omega \setminus E \end{array} \right.$$ This act assigns outcomes prescribed by $f$ for all states in the event $E$ and outcomes prescribed by $g$ for all states not in $E$.

Event $E \in \Sigma_{\Omega}$ is simple $\succeq$-\emph{null} if for all simple acts $f, g, h \in \mathcal{F}$, $$f_E h \succeq g_E h.$$ Otherwise, $E$ is simple $\succeq$-\emph{non-null}. A null event is an event for which the decision maker does not care which outcomes arises.

Below Assumption~\ref{assSEU} implies that the \emph{Sure Thing Principle} holds. Applied to events representing partitions, it implies that for any $T \geq 1$, $\bm{\pi}^T \in \Pi^T$, and any acts $f, g, h, h' \in \mathcal{F}$,
$$f_{[\bm{\pi}^T]} h \succeq g_{[\bm{\pi}^T]} h \ \mbox{ if and only if } \  f_{[\bm{\pi}^T]} h' \succeq g_{[\bm{\pi}^T]} h'.$$ That is, preferences over acts are just affected by the states for which outcomes differ among acts. Given that the Sure Thing Principle holds for events representing partitions, we can define \emph{partition-conditional preferences} as follows: For any $f, g \in \mathcal{F}$, $T \geq 1$, and $\bm{\pi}^T \in \Pi^T$, $$f \succeq_{\bm{\pi}^T} g \mbox{ if and only if } f_{[\bm{\pi}^T]} h \succeq g_{[\bm{\pi}^T]} h$$ for some $h \in \mathcal{F}$. By the Sure Thing Principle for partition events, it immediately follows that $f_{[\bm{\pi}^T]} h \succeq g_{[\bm{\pi}^T]} h$ for all $h \in \mathcal{F}$. Note that conditional on $\bm{\pi}^T$, we trivially have that the event $\Omega \setminus [\bm{\pi}^T]$ is simple $\succeq_{\bm{\pi}^T}$-null. That is, for any simple acts $f, g, h \in \mathcal{F}$, $f_{\Omega \setminus [\bm{\pi}^T]} h \succeq_{\bm{\pi}^T} g_{\Omega \setminus [\bm{\pi}^T]} h$. Thus, by the Sure Thing Principle, the definition of partition-conditional preferences makes sense.

\begin{assumption}[Partition Dependent Countable-Additive Subjective Expected Utility]\label{assSEU} We assume that $\succeq_{\bm{\pi}^T}$, $\bm{\pi}^T \in \Pi^T$, $T \geq 1$, admits a countable-additive Subjective Expected Utility representation. That is, for all $f, g \in \mathcal{F}$, $$f \succeq_{\bm{\pi}^T} g$$ if and only if
\begin{eqnarray*} \int_{\Omega} u \circ f \ d \mu(\cdot \mid [\bm{\pi}^T]) & \geq & \int_{\Omega} u \circ g \ d \mu(\cdot \mid [\bm{\pi}^T]).
\end{eqnarray*}
\normalsize for a continuous utility function $u: \mathbb{R} \longrightarrow \mathbb{R}$ and a countable additive probability measure $\mu \in \Delta(\Omega)$.

If $\Omega$ is simple $\succeq_{\bm{\pi}^T}$-null, then $\mu(\cdot \mid [\bm{\pi}^T])$ is arbitrary and $u$ is constant.

If $\Omega$ is simple $\succeq_{\bm{\pi}^T}$-non-null but no two disjoint simple $\succeq_{\bm{\pi}^T}$-non-null events exist in $\Sigma_{\Omega}$, then $\mu(\cdot \mid [\bm{\pi}^T])$ assigns probability $1$ to every simple $\succeq_{\bm{\pi}^T}$-non-null event and probability $0$ to every simple $\succeq_{\bm{\pi}^T}$-null event, and $u$ is unique up to continuous strictly increasing transformations.

Otherwise, the utility function is unique up to positive affine transformations and $\mu(\cdot \mid [\bm{\pi}^T])$ is uniquely determined.
\end{assumption}

In order not to distract from the focus of our work on the behavioral characterization of prediction rules, we state the representation here just as an assumption and relegate its characterization in terms of properties imposed on partition conditional preferences to Appendix~\ref{Wakker}. It is an immediate corollary of a representation theorem of subjective expected utility with countable additive probabilities measures due to Wakker (1989, Chapter V.5). Note that the utility function $u$ is assumed to be independent of the partition, an assumption that is also characterized in Appendix~\ref{Wakker}.

Note that $\Omega = [\{\{1\}\}]$. That is, in any state it is true that the partition of the first period is the singleton partition consisting of the set that contains period one. Thus, $\succeq \ = \ \succeq_{\{\{1\}\}}$. We let $\mu$ denote the unconditional probability measure associated with the representation of the unconditional preference $\succeq$.

\section{Partition Exchangeability\label{SPE}}

For any $T, T' \geq 1$ with $T > T'$ and partitions $\bm{\pi}^T \in \Pi^T$, $\bm{\pi}^{T'} \in \Pi^{T'}$ with $\bm{\pi}^T_{|T'} = \bm{\pi}^{T'}$, we have $\left[\bm{\pi}^T_{|T'}\right] = \left\{\omega \in \Omega : \pi^T(\omega)_{|T'} = \bm{\pi}^T_{|T'}\right\} = \left\{\omega \in \Omega : \pi^{T'}(\omega) = \bm{\pi}^{T'}\right\} = \left[\bm{\pi}^{T'}\right]$ by condition (i) of the mappings $\left\{\pi^T\right\}_{T = 1, ...}$. By Assumption~\ref{assSEU}, $\mu\left(\left[\bm{\pi}^T_{|T'}\right]\right) = \mu\left(\left[\bm{\pi}^{T'}\right]\right)$.

For any $T \geq 1$ and $\bm{\pi}^T \in \Pi^T$, let $a^T(\bm{\pi}^T) = (a_1^T(\bm{\pi}^T), ..., a_T^T(\bm{\pi}^T))$ be the \emph{partition vector} of $\bm{\pi}^T$ defined by $a^T_i(\bm{\pi}^T)$ being the number of partition cells of $\bm{\pi}^T$ with cardinality $i$. We emphasize that $a_i^T(\bm{\pi}^T)$ is \emph{not} the cardinality of some partition cell but the number of cells of partition $\bm{\pi}^T$ with cardinality $i$. We can understand vector $a^T(\bm{\pi}^T)$ as a ``frequency of frequencies'' because it lists the frequency of the frequencies of distinct objects sampled up to $T$, i.e., how many different objects have this or that frequency. 

Suppose now that the decision maker draws objects one after another and the sampling procedure does not change from one draw to another. By the virtue of not being able to distinguish between different novel objects before they are sampled, the decision maker should be ex ante indifferent between betting on let's say the partitions of sampling times $\{\{1, 2\}, \{3\}\}$ and $\{\{1\}, \{2, 3\}\}$. Note that both partitions have the same partition vector. That is, both partitions have exactly one cell of cardinality one and one of cardinality two. The following property models this idea of symmetry of the situation.

\begin{property}[Partition Symmetry]\label{PE} For any $T > 1$ and partitions $\bm{\pi}^T, \bm{\tilde{\pi}}^T \in \Pi^T$ with $a^T(\bm{\pi}^T) = a^T(\bm{\tilde{\pi}}^T)$,
$$1_{\left[\bm{\pi}^T\right]} 0 \sim z \ \mbox{ if and only if } \ 1_{\left[\bm{\tilde{\pi}}^T\right]} 0 \sim z$$ for some $z \in \mathbb{R}$. 
\end{property}
In words, Property~\ref{PE} states that the decision maker would be indifferent between bets on different partitions with the same partition vector. Here, $1_{\left[\bm{\pi}^T\right]} 0$ is the act that pays $1$ in the event $\left[\bm{\pi}^T\right]$ and zero otherwise.\footnote{It should be clear that by Assumption~\ref{assSEU}, we have a well-defined certainty equivalent and so $1_{\left[\bm{\pi}^T\right]} 0 \sim z$ if and only if $1_{\left[\bm{\tilde{\pi}}^T\right]} 0 \sim z$ implies $1_{\left[\bm{\pi}^T\right]} 0 \sim 1_{\left[\bm{\tilde{\pi}}^T\right]} 0$. However, since one of our objectives is to state testable properties of the theory, we find our more transparent statements of the properties more useful.}  

Property~\ref{PE} characterizes partition exchangeability, the central property of the literature on exchangeable random partitions in statistics.

\begin{proposition}\label{PE_result} Let $\left\{\succeq_{\bm{\pi}^T}\right\}_{\bm{\pi}^T \in \Pi^T, T \geq 1}$ satisfy Assumption~\ref{assSEU} and denote by $\left(\mu(\cdot \mid [\bm{\pi}^T])\right)_{\bm{\pi}^T \in \Pi^T, T \geq 1}$ the associated partition-conditional probability measures. Preferences $\left\{\succeq_{\bm{\pi}^T}\right\}_{\bm{\pi}^T \in \Pi^T, T \geq 1}$ satisfy Property~\ref{PE} if and only if for any $T \geq 1$,
\begin{itemize}
\item[(i)] Partition Exchangeability: for any partitions $\bm{\pi}^T, \bm{\tilde{\pi}}^T \in \Pi^T$ with $a^T(\bm{\pi}^T) = a^T(\bm{\tilde{\pi}}^T)$, $$\mu\left(\left[\bm{\pi}^T\right]\right) = \mu\left(\left[\bm{\tilde{\pi}}^T\right]\right).$$

\item[(ii)] for any partition $\bm{\pi}^T \in \Pi^T$
    $$\mu\left(\left[\bm{\pi}^{T+1}\right] \mid \left[\bm{\pi}^T\right]\right) = \mu\left(\left[\bm{\tilde{\pi}}^{T+1}\right] \mid \left[\bm{\pi}^T\right]\right)$$ for $\bm{\pi}^{T+1}, \bm{\tilde{\pi}}^{T+1} \in \Pi^{T+1}$ with $\bm{\pi}^{T+1}_{|T} = \bm{\tilde{\pi}}^{T+1}_{|T} = \bm{\pi}^T$ and $a^{T+1}(\bm{\pi}^{T+1}) = a^{T+1}(\bm{\tilde{\pi}}^{T+1})$.
\end{itemize}
\end{proposition}

The proof is contained in Appendix~\ref{proofs}.

The decision maker's subjective beliefs are such that any two partitions with the same partition vector must be assigned the same subjective probability. Similarly, conditional on a partition $\bm{\pi}^T$, any partitions of sampling times $\{1, ..., T+1\}$ that are consistent with $\bm{\pi}^T$ must get the same subjective probability if they give rise to the same partition vector.

\begin{property}\label{full_support} For any $T$ and $\bm{\pi}^T \in \Pi^T$, $[\bm{\pi}^T]$ is simple $\succeq$-non-null.
\end{property}

This property implies that ex ante for any $T \geq 1$, the decision maker does not rule out any $\bm{\pi}^T \in \Pi^T$ from arising. This is a cautiousness or admissibility property. 

For any $T \geq 1$ and $k \in \{1, ..., T\}$, define the event $$\left[|\pi^{T}(\cdot)(T)| = k\right] := \left\{\omega \in \Omega : | \pi^{T}(\omega)(T)| = k\right\}.$$ This is the event that the object drawn in $T$ has been drawn the $k$th time. (Recall that for any partition $\bm{\pi}^T$, we denote by $\bm{\pi}^T(t)$ the partition cell containing period $t$.) Clearly, since $\pi^T$ is measurable for all $T \geq 1$, we have $\left[|\pi^{T}(\cdot)(T)| = k\right] \in \Sigma_{\Omega}$ for any $T \geq 1$ and $k \leq T$.

Define $$\Pi^{T, k} := \left\{\bm{\pi}^T \in \Pi^T : |\bm{\pi}^T(t)| = k, \mbox{ for some } t \in \{1, ..., T\} \right\}.$$ This is the set of partitions of $\{1, ..., T\}$ that have a partition cell of cardinality $k$.

\begin{property}[Frequency Dependence of a Known]\label{frequency_dependence} For all $T \geq 1$, $k \in \{1, ...T\}$, $\bm{\pi}^T, \bm{\tilde{\pi}}^T \in \Pi^{T, k}$,
$$1_{\left[|\pi^{T+1}(\cdot)(T+1)| = k + 1\right]} 0 \sim_{\bm{\pi}^T} z \ \mbox{ if and only if } \ 1_{\left[|\pi^{T+1}(\cdot)(T+1)| = k + 1\right]} 0 \sim_{\bm{\tilde{\pi}}^T} z$$ for some $z \in \mathbb{R}$. 
\end{property}

The property means the decision maker's evaluation of the event that the object sampled next is an object that has been previously sampled $k$ times is invariant to the partition as long as it contains a partition cell of cardinality $k$. Thus, the evaluation of sampling a known particular object depends only on the frequency with which it has been sampled before and on the sampling time.

\begin{property}[Frequency Dependence of Novelty]\label{frequency_dependence_new} For all $T \geq 1$ and $\bm{\pi}^T, \bm{\tilde{\pi}}^T \in \Pi^T$ with $|\bm{\pi}^T | = |\bm{\tilde{\pi}}^T|$,
$$1_{\left[|\pi^{T+1}(\cdot)(T+1)| = 1\right]} 0 \sim_{\bm{\pi}^T} z \ \mbox{ if and only if } \ 1_{\left[|\pi^{T+1}(\cdot)(T+1)| = 1\right]} 0 \sim_{\bm{\tilde{\pi}}^T} z$$ for some $z \in \mathbb{R}$. 
\end{property}

This property pertains to the evaluation of sampling a novel object. When $|\bm{\pi}^{T+1}(T+1)| = 1$, then the partition cell containing sampling time $T+1$ is a singleton. That is, at $T+1$ an object is sampled that has never been sampled before. Note that Property~\ref{frequency_dependence_new} is not a special case of Property~\ref{frequency_dependence} because Property~\ref{frequency_dependence_new} pertains to any two partitions with the same number of partition cells while Property~\ref{frequency_dependence} pertains to any two partitions with a partition cell of cardinality $k$. The number of partition cells of a partition represents the number of times novelty has been sampled. Thus, Property~\ref{frequency_dependence_new} says that the evaluation of a novel object depends only on how often a novel object has been sampled in the past and on the sampling time.

Define the event $$\left[|\pi^{T+1}(\cdot)| = |\pi^T(\cdot)| + 1\right] := \left\{\omega \in \Omega : |\pi^{T+1}(\omega)| = |\pi^T(\omega)| + 1\right\}.$$ Since $\pi_i^T$ is measurable, $\left[|\pi^{T+1}(\cdot)| = |\pi^T(\cdot)| + 1\right] \in \Sigma_{\Omega}$. To interpret this event recall that when a novel object is sampled in period $T$, a new partition cell is added beyond the partition cells already in partition of prior sampling times. Thus, this is the event that a novel object is sampled at $T+1$. Note that $\left[|\pi^{T+1}(\cdot)| = |\pi^T(\cdot)| + 1\right] = \left[|\pi^{T+1}(\cdot)(T+1)| = 1\right]$. That is, when the next object adds a new partition cell then the partition cell of the sampling times of objects corresponding to the next object must be a singleton and vice versa. These are just different statement of the same event that we will use whenever convenient.

Also define the event
$$\left[|\pi_j^{T+1}(\cdot)| = |\pi^T_j(\cdot)| + 1\right] := \left\{\omega \in \Omega : |\pi_j^{T+1}(\omega)| = |\pi_j^{T}(\omega)| + 1\right\}.$$ Clearly, since $\pi^T$ is measurable, we have $\left[|\pi_j^{T+1}(\cdot)| = |\pi^T_j(\cdot)| + 1\right] \in \Sigma_{\Omega}$ for any $j$. To interpret the event, recall that when a previously sampled object is sampled again in $T+1$, it must extend one partition cell of prior sampling times by exactly one element. Thus, this is the event that the object sampled previously at sampling times in partition cell $\bm{\pi}_j^T$ is sampled again in period $T+1$. Note the difference between events $\left[|\pi_j^{T+1}(\cdot)| = |\pi_j^T(\cdot)| + 1\right]$ and $\left[|\pi^{T+1}(\cdot)| = |\pi^T(\cdot)| + 1\right]$. Former event pertains to the cardinality of the $j$th partition cell (in the order of appearance). Latter event pertains to the cardinality of partitions, i.e., the number of partition cells.

The following theorem behaviorally characterizes the two-parameter model and prediction rule of Pitman (1995) and Zabell (1997).

\begin{theorem}\label{Zabell_SEU} Let $\left\{\succeq_{\bm{\pi}^T}\right\}_{\bm{\pi}^T \in \Pi^T, T \geq 1}$ satisfy Assumption~\ref{assSEU} and denote by $\left(\mu(\cdot \mid [\bm{\pi}^T])\right)_{\bm{\pi}^T \in \Pi^T, T \geq 1}$ the partition-conditional probability measures and $u$ the utility function of the representation. The collection of partition-dependent preferences $\{\succeq_{\bm{\pi}^T}\}_{\bm{\pi}^T \in \Pi^T, T \geq 1}$ satisfy Properties~\ref{PE} to~\ref{frequency_dependence_new} if and only if for any $T \geq 1$, $\bm{\pi}^T \in \Pi^{T}$, and $j$,
    \begin{eqnarray} \mu\left( \left[|\pi_j^{T+1}(\cdot)| = |\pi_j^T(\cdot)| + 1\right] \mid \left[\bm{\pi}^T\right] \right) & = &
    \frac{|\bm{\pi}^T_j| - \alpha}{T + \theta} \label{known} \\
    \mu\left( \left[|\pi^{T+1}(\cdot)| = |\pi^T(\cdot)| + 1\right] \mid \left[\bm{\pi}^T\right] \right) & = &  \frac{\alpha |\bm{\pi}^T| + \theta}{T + \theta}. \label{novel}
\end{eqnarray} for $\alpha \in [0, 1)$ and $\theta > - \alpha$. Moreover, let $z \in \mathbb{R}$ be defined by the choice behavior $$1_{\left[\{\{1\}, \{2, 3\}, \{4\}\}\right]} 0 \sim_{\{\{1\}, \{2, 3\}\}} z$$ and $k \in \mathbb{R}$ be defined by the choice behavior $$1_{\left[\{\{1, 4\}, \{2, 3\}\}\right]} 0 \sim_{\{\{1\}, \{2, 3\}\}} k.$$ If $u$ is normalized such that $u(0) = 0$, then
\begin{eqnarray} \alpha & = & \frac{3 u(k) + u(z) - u(1)}{2 u(k) + u(z) - u(1)} \label{alpha}\\
\theta & = & \frac{2 u(1) - 6 u(k) - 3 u(z)}{2 u(k) + u(z) - u(1)}. \label{theta}
\end{eqnarray}
\end{theorem}

The proof is contained in Appendix~\ref{proofs}. It extends results by Zabell (1997) to a behavioral characterization. Equations~(\ref{known}) and~(\ref{novel}) correspond to the predictive probabilities for sampling a particular known object and sampling a novel object, respectively, as informally introduced in the Introduction. Observe that in equation~(\ref{known}) we consider the cardinality of the partition \emph{cell} $\bm{\pi}^T_j$ while in equation~(\ref{novel}) we consider the cardinality of the partition $\bm{\pi}^T$, i.e., the number of partition cells of the partition. Observe further that the sum of probabilities (over $j$) in equations~(\ref{known}) plus the probability in equation~(\ref{novel}) add up to one. While parameters $\alpha$ and $\theta$ are exogenously given in the two-parameter Pitman-Zabell model, casting it in terms of subjective probability theory allows us to derive these parameters endogenously from choice behavior. We can elicit the conditional certainty equivalents $z$ and $k$ via $1_{\left[\{\{1\}, \{2, 3\}, \{4\}\}\right]} 0 \sim_{\{\{1\}, \{2, 3\}\}} z$ and $1_{\left[\{\{1, 4\}, \{2, 3\}\}\right]} 0 \sim_{\{\{1\}, \{2, 3\}\}} k$, respectively. By Assumption~\ref{assSEU}, we have a subjective expected utility representation with continuous utility function $u$ unique up to affine transformations. This allows us to normalize $u(0) = 0$. (Note that the utility index $u$ is not exogenous given but can in principle be endogenous derived from choices; see Appendix~\ref{Wakker}). Using this utility index and the particular form of subjective beliefs implied by Properties~\ref{PE} to~\ref{frequency_dependence_new}, we can recast above choices in terms of the subjective expected utility representation, and derive two equations whose solution give us $\alpha$ and $\theta$, respectively (see the end of the Proof of Theorem~\ref{Zabell_SEU}). 

For the next result, we need some notation: For $t = 1, 2, ...$, $x, y \in \mathbb{R}$, let $(x)_{t \uparrow y}$ denote the \emph{$t$-th factorial power of $x$ with increment $y$}, i.e.,
$$(x)_{t \uparrow y} := \prod_{i = 0}^{t-1} (x + i y).$$

Combining Theorem~\ref{Zabell_SEU} with existing results in the literature (i.e., Pitman, 1995, 2006) yields the following properties of beliefs of a subjective expected utility maximizer who satisfies Properties~\ref{PE} to~\ref{frequency_dependence_new}:

\begin{corollary} Let $\mu$ be the subjective probability measure of the representation of the collection of partition-dependent preferences $\{\succeq_{\bm{\pi}^T}\}_{\bm{\pi}^T \in \Pi^T, T \geq 1}$ satisfying Assumption~\ref{assSEU} and Properties~\ref{PE} to~\ref{frequency_dependence_new}. For $T \geq 1$,
\begin{itemize}
\item[(i)] Subjective belief in partition $\bm{\pi}^T = (\bm{\pi}^T_1, ..., \bm{\pi}^T_k) \in \Pi^T$ (some $k \leq T$):
    \begin{eqnarray} \mu\left(\left[\bm{\pi}^T\right]\right) & = & \frac{(\theta + \alpha)_{k-1 \uparrow \alpha}}{(\theta + 1)_{T - 1 \uparrow 1}} \prod_{j = 1}^k (1 - \alpha)_{|\bm{\pi}^T_j| - 1 \uparrow 1} \label{EPF_two_parameters}
    \end{eqnarray}
\item[(ii)] Subjective probability of $k$ novelties by sampling time $T$:
    \begin{eqnarray} \mu\left(\left[|\pi^T(\cdot)| = k\right]\right) & = & \frac{(\theta - \alpha)_{k-1 \uparrow \alpha}}{(\theta + 1)_{T - 1 \uparrow 1}} S_{\alpha}(T, k)
    \end{eqnarray} where $S_{\alpha}(T, k)$ is the generalized Stirling number and $\left[|\pi^T(\cdot)| = k\right] := \left\{\omega \in \Omega: |\pi^T(\omega)| = k\right\} \in \Sigma_{\Omega}$.
\item[(iii)] Expected number of novelties by sampling time $T$:
\begin{eqnarray} \mathbb{E}\left[|\pi^T(\cdot)|\right] & = & \sum_{i = 1}^T \frac{(\theta + \alpha)_{i - 1 \uparrow 1}}{(\theta + 1)_{i - 1 \uparrow 1}}
    \end{eqnarray}
\end{itemize} where $\alpha$ and $\theta$ are given by equations~(\ref{alpha}) and~(\ref{theta}), respectively.  
\end{corollary}

For (i), see Pitman (1995, Proposition 9). For (ii) and (iii), see Pitman (2006, p. 66). For generalized Stirling numbers, see Pitman (2006, p. 20-21). The ``closed-form'' expressions in previous corollary should prove to be useful in applications. 

Next, we consider a strengthening of Property~\ref{frequency_dependence_new}:

\begin{property}[Sampling Time Dependence of Novelty]\label{frequency_independence} For all $T \geq 1$ and $\bm{\pi}^T, \bm{\tilde{\pi}}^T \in \Pi^T$,
$$1_{\left[|\pi^{T+1}(\cdot)(T+1)| = 1\right]} 0 \sim_{\bm{\pi}^T} z \mbox{ if and only if }  1_{\left[|\tilde{\pi}^{T+1}(\cdot)(T+1)| = 1\right]} 0 \sim_{\bm{\tilde{\pi}}^T} z$$ for some $z \in \mathbb{R}$. 
\end{property}

Like Property~\ref{frequency_dependence_new}, this property pertains to the evaluation of sampling novelty. Recall that when $|\bm{\pi}^{T+1}(T+1)| = 1$, then the partition cell containing sampling time $T+1$ is a singleton. It must be the case that at $T+1$ an object is sampled that has never been sampled before. Yet, while Property~\ref{frequency_dependence_new} depends on the number of partition cells, Property~\ref{frequency_independence} is independent of it. The number of partition cells of a partition represents the number of times novel objects have been sampled. Thus, Property~\ref{frequency_independence} says that the evaluation of a novel object is independent of how often a novelty has occurred in the past and just depends on the sampling time.

The following theorem characterizes the prediction rule associated with the famous one parameter model known as the Ewens' sampling formula. The representation is a special case of the representation in Theorem~\ref{Zabell_SEU} for $\alpha = 0$.

\begin{theorem}\label{Ewens_SEU} Let $\left\{\succeq_{\bm{\pi}^T}\right\}_{\bm{\pi}^T \in \Pi^T, T \geq 1}$ satisfy Assumption~\ref{assSEU} and denote by $\left(\mu\left(\cdot \mid \left[\bm{\pi}^T\right]\right)\right)_{\bm{\pi}^T \in \Pi^T, T \geq 1}$ the partition-conditional probability measures of the representation. The collection of partition-dependent preferences $\{\succeq_{\bm{\pi}^T}\}_{\bm{\pi}^T \in \Pi^T, T \geq 1}$ satisfy Properties~\ref{PE} to~\ref{frequency_dependence} and~\ref{frequency_independence} if and only if for any $T \geq 1$, $\bm{\pi}^T \in \Pi^{T}$, and $j$,
    \begin{eqnarray} \mu\left( \left[|\pi_j^{T+1}(\cdot)| = |\pi_j^T(\cdot)| + 1\right] \mid \left[\bm{\pi}^T\right] \right) & = &
    \frac{|\bm{\pi}^T_j|}{T + \theta} \label{known1} \\
    \mu\left(\left[|\pi^{T+1}(\cdot)| = |\pi^T(\cdot)| + 1\right] \mid \left[\bm{\pi}^T\right]\right) & = &  \frac{\theta}{T + \theta}. \label{novel1}
\end{eqnarray}
for $\theta > 0$. Moreover, let $z \in \mathbb{R}$ be defined by the choice behavior $1_{\left[\{\{1\}, \{2\}\}\right]} 0 \sim_{\{\{1\}\}} z.$ If $u$ is normalized such that $u(0) = 0$, then
\begin{eqnarray} \theta & = & \frac{u(z)}{u(1) - u(z)}. \label{theta1}
\end{eqnarray}
\end{theorem}

The proof follows by arguments analogous to the proof of Theorem~\ref{Zabell_SEU}. Instead of Property~\ref{frequency_dependence_new} we use Property~\ref{frequency_independence}. The derived partition-conditional probability of a novel object depends only on the sample size $T$. By Zabell (1997, Corollary p. 268), this implies the prediction rule of the one-parameter model. The converse is straightforward like in the proof of Theorem~\ref{Zabell_SEU}. The parameter $\theta$ can be identified from choices. This follows from arguments similar to the ones in the last part of the proof of Theorem~\ref{Zabell_SEU}. Equation~(\ref{known1}) is the predictive probability of sampling a known object and equation~(\ref{novel1}) is the predictive probability of sampling a novel object. Again, we emphasize that while parameter $\theta$ is exogenously given in the theory of Ewens' sampling formula, the parameter can be now endogenously derived from the choice $1_{\left[\{\{1\}, \{2\}\}\right]} 0 \sim_{\{\{1\}\}} z$ and the subjective expected utility representation under Assumption~\ref{assSEU} as well as Properties~\ref{PE} to~\ref{frequency_dependence} and~\ref{frequency_independence} using equation~(\ref{theta1}).   

The major difference between beliefs of Theorems~\ref{Zabell_SEU} and~\ref{Ewens_SEU} is the beliefs about becoming aware of a novel object depend on awareness of objects (i.e., the number of objects) in Theorem~\ref{Zabell_SEU} but are independent of awareness of objects in Theorem~\ref{Ewens_SEU}. For the one-parameter model of Theorem~\ref{Ewens_SEU}, we can think of the decision maker as keeping a ``mental'' urn in which the black ball has initial weight $\theta$. Whenever a colored ball is drawn, it is returned to the urn together with another ball of the same color. Whenever a black ball is drawn, it is returned to the urn together with a ball of a new color.

Theorem~\ref{Ewens_SEU}, Ewens (1972), and Hoppe (1984) imply now immediately
\begin{corollary} Let $\mu$ be the subjective probability measure of the representation of the collection of partition-dependent preferences $\{\succeq_{\bm{\pi}^T}\}_{\bm{\pi}^T \in \Pi^T, T \geq 1}$ satisfying Assumption~\ref{assSEU}, Properties~\ref{PE} to~\ref{frequency_dependence}, and Property~\ref{frequency_independence}. Then for any $T \geq 1$,
\begin{itemize}
\item[(i)] and $\bm{\pi}^T = (\bm{\pi}^T_1, ..., \bm{\pi}^T_k) \in \Pi^T$ (some $k < T$) we have the Ewens' sampling formula
    \begin{eqnarray} \mu\left(\left[\bm{\pi}^T\right]\right) & = & \frac{\theta^k}{(\theta)_{T \uparrow 1}} \prod_{j = 1}^k (|\bm{\pi}^T_j| - 1)!
    \end{eqnarray}
\item[(ii)] Probability of $k$ novelties up to sampling time $T$:
\begin{eqnarray} \mu\left(\left[|\pi^T(\cdot)| = k\right]\right) & = & \frac{\theta^k}{(\theta)_{T \uparrow 1}} s_{T, k}
\end{eqnarray} where $s_{T, k}$ is the $(T, k)$-Stirling number of the first kind.
\item[(iii)] Expected number of novelties by sampling time $T$:
\begin{eqnarray} \mathbb{E}\left[|\pi^T(\cdot)|\right] & = & \sum_{i = 1}^T \frac{\theta}{\theta + i - 1}
\end{eqnarray}
\end{itemize}
with $\theta$ is given by equation~(\ref{theta1}). 
\end{corollary}

Part (i) follows from Theorem~\ref{Ewens_SEU} and Ewens (1972). Theorem~\ref{Ewens_SEU} implies part (ii), see for instance Crane (2016, p. 2). Part (iii) follows from Theorem~\ref{Ewens_SEU} and Pitman (2006, p. 66). Again, the closed form expressions should prove to be useful in applications. 


A special case of the one-parameter prediction rule of Theorem~\ref{Ewens_SEU} with $\theta = 1$ is the \emph{De Morgan prediction rule}, probably the oldest prediction rule for sampling with novelties (De Morgan, 1838, 1845). See Zabell (1997, 1992) for discussions. It is given by for any $T \geq 1$, $\bm{\pi}^T \in \Pi^{T}$, and $j$,
\begin{eqnarray} \mu\left( \left[|\pi_j^{T+1}(\cdot)| = |\pi_j^T(\cdot)| + 1\right] \mid \left[\bm{\pi}^T\right] \right) & = &
    \frac{|\bm{\pi}^T_j|}{T + 1} \label{knownDM} \\
    \mu\left(\left[|\pi^{T+1}(\cdot)| = |\pi^T(\cdot)| + 1\right] \mid \left[\bm{\pi}^T\right]\right) & = &  \frac{1}{T + 1}. \label{novelDM}
\end{eqnarray} Equation~(\ref{knownDM}) is the predictive probability of sampling a known object and equation~(\ref{novelDM}) is the predictive probability of sampling a novel object. This prediction rule can be interpreted as the decision maker keeping a ``mental'' urn in which there is initially one black ball. If a colored ball it drawn, it is returned to the urn together with a ball of the same color. If the black ball is drawn, it is returned together with a ball of a new color. The De Morgan model arises under Assumption~\ref{assSEU} when choices of the decision maker satisfy Properties~\ref{PE} to~\ref{frequency_dependence} and Property~\ref{frequency_independence} and $\frac{u(z)}{u(1)} = \frac{1}{2}$ where $z \in \mathbb{R}$ defined by $1_{\left[\{\{1\}, \{2\}\}\right]} 0 \sim_{\{\{1\}\}} z$.  

In choice experiments, stakes are often small and researchers may be willing to assume risk neutrality as an additional convenient identifying assumption. Risk neutrality implies that the utility function $u$ is linear. Under risk neutrality, there are particular straightforward behavioral implications of the three models. Denote by $T_{[\bullet]}0$ the act that pays $T$ when a novel object is drawn in period $T$ and zero otherwise. Similar, denote by $T_{[j]} 0$ the act that pays $T$ when the $j$th known object is encountered in period $T$ and zero otherwise.

\begin{proposition}\label{risk_neutrality} Assume risk neutrality, i.e., $u$ is linear. Under the previous assumptions, respectively, the decision maker's subjective beliefs follow
\begin{enumerate}
\item the De Morgan prediction rule (i.e., Theorem~\ref{Ewens_SEU} with $\theta = 1$) if
\begin{eqnarray*} (T+1)_{[\bullet]} 0 & \sim_{\bm{\pi}^T} & 1 \\
(T+1)_{[j]} 0 & \sim_{\bm{\pi}^T} & |\bm{\pi}_j^T|
\end{eqnarray*}
\item the prediction rule of the Ewens model (i.e., Theorem~\ref{Ewens_SEU}) if for some $\theta > 0$,
\begin{eqnarray*} (T+\theta)_{[\bullet]} 0 & \sim_{\bm{\pi}^T} & \theta \\
(T+\theta)_{[j]} 0 & \sim_{\bm{\pi}^T} & |\bm{\pi}_j^T|
\end{eqnarray*}
\item the prediction rule of the two-parameter model of Pitman (1995) and Zabell (1997) (i.e., Theorem~\ref{Zabell_SEU}) if for some $\alpha \in [0, 1)$ and $\theta > -\alpha$,
\begin{eqnarray*} (T+\theta)_{[\bullet]} 0 & \sim_{\bm{\pi}^T} & (\alpha | \bm{\pi}^T | + \theta) \\
(T+\theta)_{[j]} 0 & \sim_{\bm{\pi}^T} & (|\bm{\pi}_j^T| - \alpha)
\end{eqnarray*}
\end{enumerate} for any $\bm{\pi}^T \in \Pi^T$ and $T \geq 1$.
\end{proposition}

The proof is contained in Appendix~\ref{proofs}.

\section{``Reverse'' Bayesianism\label{RB}}

Karni and Vier{\o} (2013) and Hayashi (2012) characterized decision theoretically changes of beliefs when becoming aware \`{a} la ``reverse'' Bayesianism. Translated to our setting, ``reverse'' Bayesianism implies that upon sampling and hence becoming aware of a novel object, relative likelihoods for previously sampled objects remain unchanged. Karni and Vier{\o} (2013) and Hayashi (2012) did not consider awareness of unawareness, i.e., beliefs about becoming aware. However, Karni and Vier{\o} (2017) and Vier{\o} (2021, 2023) extended the approach to awareness of unawareness. Yet, they consider the case of becoming aware of another outcome while we consider the case of becoming aware of a state/object.

For $T \geq 1$ and $i \leq T$, define events $$\left[T \in \pi_i^T(\cdot)\right] := \left\{\omega \in \Omega : T \in \pi_i^T(\omega)\right\} \in \Sigma_{\Omega}.$$ This is the event that the object sampled in period $T$ is the same as sampled at times in the $i$th partition cell (in the order of appearance) of some partition up to sampling time $T$.

\begin{property}[Consistency Upon Sampling Novelty]\label{novelty_sampling_consistency} $\{\succeq_{\bm{\pi}^T}\}_{\bm{\pi}^T \in \Pi^T, T = 1, ...}$ satisfy consistency upon sampling novelty if for any $T$, $\bm{\pi}^T \in \Pi^T$, and $i, j \leq | \bm{\pi}^T|$, there are $x, y \in \mathbb{R}$ such that
\begin{eqnarray}\label{sampling_consistency_condition} x_{\left[T+1 \in \pi^{T+1}_i(\cdot)\right]} 0 \sim_{\bm{\pi}^T} y_{\left[T+1 \in \pi^{T+1}_j(\cdot)\right]}0 & \mbox{ iff } & x_{\left[T+2 \in \pi^{T+2}_i(\cdot)\right]} 0 \sim_{\bm{\pi}^{T+1}} y_{\left[T+2 \in \pi^{T+2}_j(\cdot)\right]}0
\end{eqnarray} for $\bm{\pi}^{T+1} = \bm{\pi}^T \cup \{T+1\}$.
\end{property}

In words, this property means that the decision maker's evaluation of sampling one known object versus another known object remains unchanged upon sampling of a novel object. Here $\bm{\pi}^{T+1} = \bm{\pi}^T \cup \{T+1\}$ means that partition $\bm{\pi}^{T+1}$ is the same as $\bm{\pi}^T$ up to sampling time $T$ and then appended by an additional partition cell $\{T+1\}$, which means a novel object is drawn in period $T+1$. The requirement $i, j \leq | \bm{\pi}^T|$ implies that partition cells $\bm{\pi}^{T+1}_i$ and $\bm{\pi}^{T+1}_j$ are not partition cells newly added in period $T+1$. That is, condition~(\ref{sampling_consistency_condition}) is about betting on that known objects are sampled in periods $T+1$ and $T+2$ rather than a novel object. 

The next property pertains to the invariance of the evaluation of sampling a known object versus another known object upon sampling a different third known object. 

\begin{property}[Consistency Upon Sampling Known Objects]\label{sampling_consistency} $\{\succeq_{\bm{\pi}^T}\}_{\bm{\pi}^T \in \Pi^T, T = 1, ...}$ satisfy consistency upon sampling known objects if for any $T$, $\bm{\pi}^T \in \Pi^T$, and $i, j \leq | \bm{\pi}^{T}|$, there are $x, y \in \mathbb{R}$ such that Condition~(\ref{sampling_consistency_condition}) holds for all $\bm{\pi}^{T+1}$ for which there exists $k \leq |\bm{\pi}^T|$ with $i \neq k \neq j$ and $\bm{\pi}^{T+1}(T+1) \cap \{1, ..., T\} = \bm{\pi}^T_k$. 
\end{property} 

The condition that there exist $k \leq |\bm{\pi}^T|$ with $i \neq k \neq j$ and $\bm{\pi}^{T+1}(T+1) \cap \{1, ..., T\} = \bm{\pi}^T_k$ means that there is a known object that (a) is sampled previously at periods in $\bm{\pi}^T_k$, (b) is not the object sampled previously at periods in $\bm{\pi}^T_i$ or $\bm{\pi}^T_j$ but (c) is sampled again in period $T + 1$. 

Both, Property~\ref{novelty_sampling_consistency} and Property~\ref{sampling_consistency}, pertain to the evaluation of known objects upon sampling of either a novel or another known object, respectively. Yet, the decision maker may also re-evaluate consistently novelty upon sampling of a known object. 

\begin{property}[Sampling Consistency for Novelty]\label{consistency_for_novelty} $\{\succeq_{\bm{\pi}^T}\}_{\bm{\pi}^T \in \Pi^T, T = 1, ...}$ satisfy sampling consistency for novelty if for any $T$, $\bm{\pi}^T \in \Pi^T$, and $i \leq | \bm{\pi}^{T}|$, there are $x, y \in \mathbb{R}$ such that
\begin{eqnarray}\label{sampling_consistency_novelty_condition} x_{\left[|\pi^{T+1}(T+1)| = 1 \right]} 0 \sim_{\bm{\pi}^T} y_{\left[T+1 \in \pi^{T+1}_i(\cdot)\right]}0 & \mbox{ iff } & x_{\left[|\pi^{T+2}(T+2)| = 1 \right]} 0 \sim_{\bm{\pi}^{T+1}} y_{\left[T+2 \in \pi^{T+2}_i(\cdot)\right]}0
\end{eqnarray}
for all $\bm{\pi}^{T+1}$ for which there exist $j \leq |\bm{\pi}^T|$ with $i \neq j$ and $\bm{\pi}^{T+1}(T+1) \cap \{1, ..., T\} = \bm{\pi}^T_j$. 
\end{property} To understand the condition, recall that $\left[|\pi^{T+1}(T+1)| = 1 \right]$ is the event that a new partition cell is added in period $T+1$, i.e., a novel object is sampled in $T+1$. Similarly, $\left[|\pi^{T+2}(T+2)| = 1 \right]$ is the event that a novel object is sampled in period $T+2$. The requirement of $\bm{\pi}^{T+1}$ for which there exist $j \leq |\bm{\pi}^T|$ with $i \neq j$ and $\bm{\pi}^{T+1}(T+1) \cap \{1, ..., T\} = \bm{\pi}^T_j$ means that a known object different from the known object sampled at periods in $\bm{\pi}^T_i$ is actually sampled in period $T+1$. Thus, Property~\ref{consistency_for_novelty} says that how sampling of novelty is evaluated relative to sampling of a known object is invariant to actually sampling a different known object. 

\begin{proposition}\label{reverse_Bayesianism} Let $\left\{\succeq_{\bm{\pi}^T}\right\}_{\bm{\pi}^T \in \Pi^T, T \geq 1}$ satisfy Assumption~\ref{assSEU} and Property~\ref{full_support}, and denote by $\left(\mu\left(\cdot \mid \left[\bm{\pi}^T\right]\right)\right)_{\bm{\pi}^T \in \Pi^T, T \geq 1}$ the associated partition-conditional probability measures. 
\begin{itemize}
\item[(i)] The collection of partition-dependent preferences $\{\succeq_{\bm{\pi}^T}\}_{\bm{\pi}^T \in \Pi^T, T \geq 1}$ satisfy Property~\ref{novelty_sampling_consistency} if and only if $\left(\mu\left(\cdot \mid \left[\bm{\pi}^T\right]\right)\right)_{\bm{\pi}^T \in \Pi^T, T \geq 1}$ satisfy ``reverse'' Bayesianism, i.e., for any $T$, $\bm{\pi}^T \in \Pi^T$, and $i, j \leq |\bm{\pi}^T|$,
\begin{eqnarray}\label{Bayesianism} \frac{\mu\left( \left[T+1 \in \pi^{T+1}_i(\cdot)\right] \mid \left[\bm{\pi}^T\right]\right)}{\mu\left(\left[T+1 \in \pi^{T+1}_j(\cdot)\right] \mid \left[\bm{\pi}^T\right]\right)} & = & \frac{\mu\left(\left[T+2 \in \pi^{T+2}_i(\cdot)\right] \mid \left[\bm{\pi}^{T+1}\right]\right)}{\mu\left(\left[T+2 \in \pi^{T+2}_j(\cdot)\right] \mid \left[\bm{\pi}^{T+1}\right]\right)}
\end{eqnarray} for $\bm{\pi}^{T+1} = \bm{\pi}^T \cup \{T+1\}$. 
\item[(ii)] They satisfy Property~\ref{sampling_consistency} if and only if $\left(\mu\left(\cdot \mid \left[\bm{\pi}^T\right]\right)\right)_{\bm{\pi}^T \in \Pi^T, T \geq 1}$ satisfy Bayesianism, i.e., for any $T$, $\bm{\pi}^T \in \Pi^T$, and $i, j \leq |\bm{\pi}^T|$, equation~(\ref{Bayesianism}) holds for all $\bm{\pi}^{T+1}$ for which there exists $k \leq |\bm{\pi}^T|$ with $i \neq k \neq j$ and $\bm{\pi}^{T+1}(T+1) \cap \{1, ..., T\} = \bm{\pi}^T_k$.
\item[(iii)] They satisfy Property~\ref{consistency_for_novelty} if and only if $\left(\mu\left(\cdot \mid \left[\bm{\pi}^T\right]\right)\right)_{\bm{\pi}^T \in \Pi^T, T \geq 1}$ satisfy ``extended'' Bayesianism, i.e., for any $T$, $\bm{\pi}^T \in \Pi^T$, and $i \leq | \bm{\pi}^{T}|$, 
\begin{eqnarray}\label{extended Bayesianism} \frac{\mu\left( \left[|\pi^{T+1}(T+1)| = 1 \right] \mid \left[\bm{\pi}^T\right]\right)}{\mu\left(\left[T+1 \in \pi^{T+1}_i(\cdot)\right] \mid \left[\bm{\pi}^T\right]\right)} & = & \frac{\mu\left(\left[|\pi^{T+2}(T+2)| = 1 \right] \mid \left[\bm{\pi}^{T+1}\right]\right)}{\mu\left(\left[T+2 \in \pi^{T+2}_i(\cdot)\right] \mid \left[\bm{\pi}^{T+1}\right]\right)}
\end{eqnarray} for all $\bm{\pi}^{T+1}$ for which there exist $j \leq |\bm{\pi}^T|$ with $i \neq j$ and $\bm{\pi}^{T+1}(T+1) \cap \{1, ..., T\} = \bm{\pi}^T_j$. 
\end{itemize}
\end{proposition}

``Reverse'' Bayesianism means that relative likelihood ratios for known objects are preserved upon sampling novel objects. Bayesianism means that the likelihood ratios for any two known alternatives are preserved upon sampling another third known object. Extended Bayesianism refers to the preservation of the relatively likelihood ratios for novelty and a known object upon sampling another known object. Property~\ref{novelty_sampling_consistency} plays the same role as  ``Awareness consistency'' in the literature on ``reverse'' Bayesianism, e.g., Karni and Vier{\o} (2013, 2017), Dominiak and Tserenjigmid (2018), and Dominiak (2022).

\begin{figure}[h!]
\caption{Partition Exchangeability Does Not Imply ``Extended'' Bayesianism\label{snapshot1}}
\begin{center}
\includegraphics[scale=0.4]{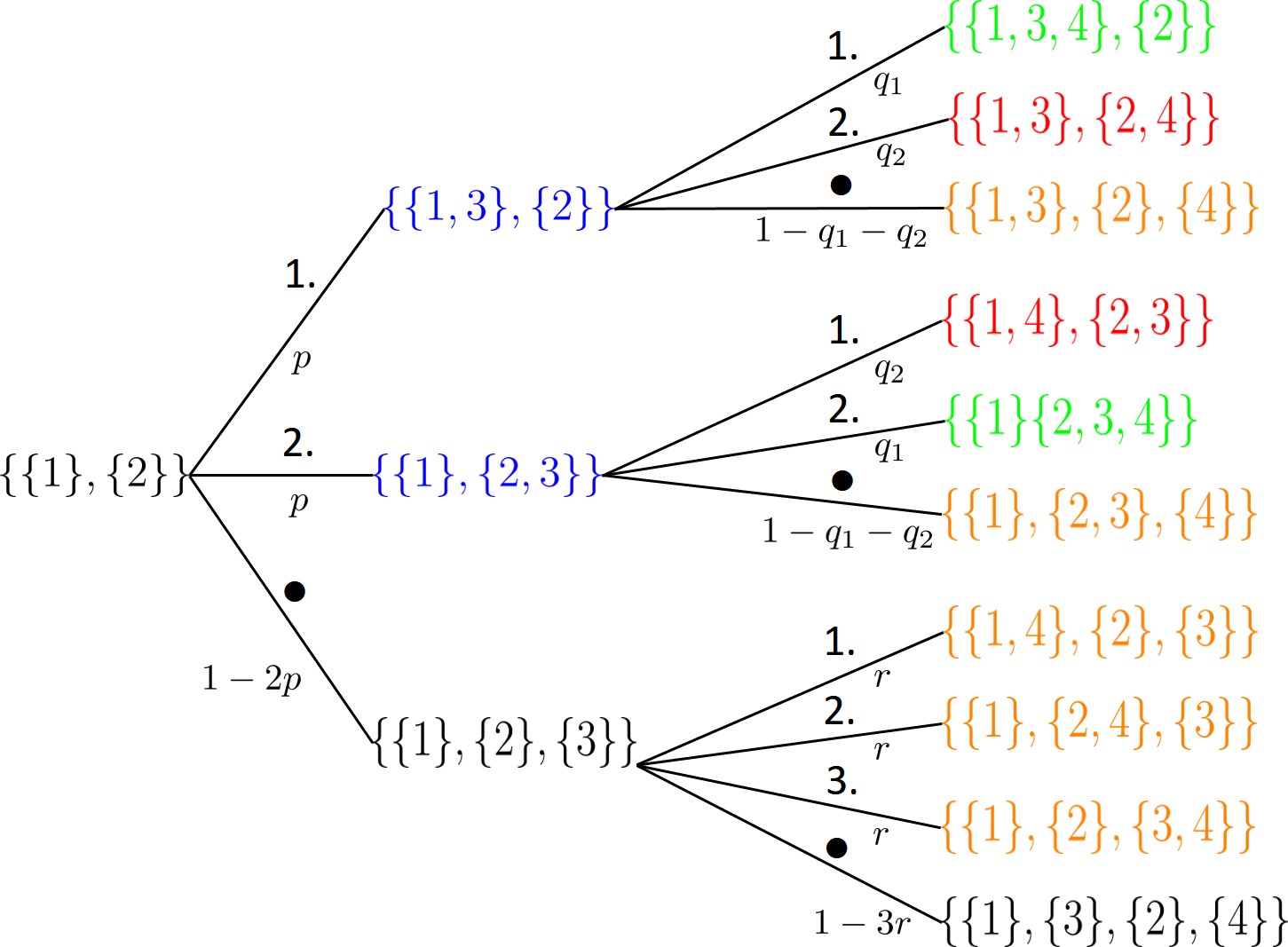}
\end{center}
\end{figure}

Proposition~\ref{reverse_Bayesianism} does not impose partition exchangeability or partition symmetry (Property~\ref{PE}). We can show that not all prediction rules consistent with partition exchangeable beliefs may satisfy ``reverse'' Bayesianism or ``extended'' Bayesianism. To see this, consider Figure~\ref{snapshot1}. It is a snapshot from Figure~\ref{T4} in the introduction focusing on the transition from partition $\{\{1\}, \{2\}\}$ in $T = 2$ to partition $\{\{1\}, \{2, 3\}\}$ in $T = 3$ via sampling again the second object. How does the ratio of probabilities of objects that are not sampled evolve? We have $\frac{\mu([1.] \mid \{\{1\}, \{2\}\})}{\mu([\bullet] \mid \{\{1\}, \{2\}\})} = \frac{p}{1-2p}$ and $\frac{\mu([1.] \mid \{\{1\}, \{2, 3\}\})}{\mu([\bullet] \mid \{\{1\}, \{2, 3\}\})} = \frac{q_2}{1-q_1-q_2}$. ``Extended'' Bayesianism requires that these ratios are equal. But partition exchangeability does not force these ratios to be equal unless further structure is imposed.

\begin{figure}[h!]
\caption{Partition Exchangeability Does Not Imply ``Reverse'' Bayesianism\label{snapshot2}}
\begin{center}
\includegraphics[scale=0.4]{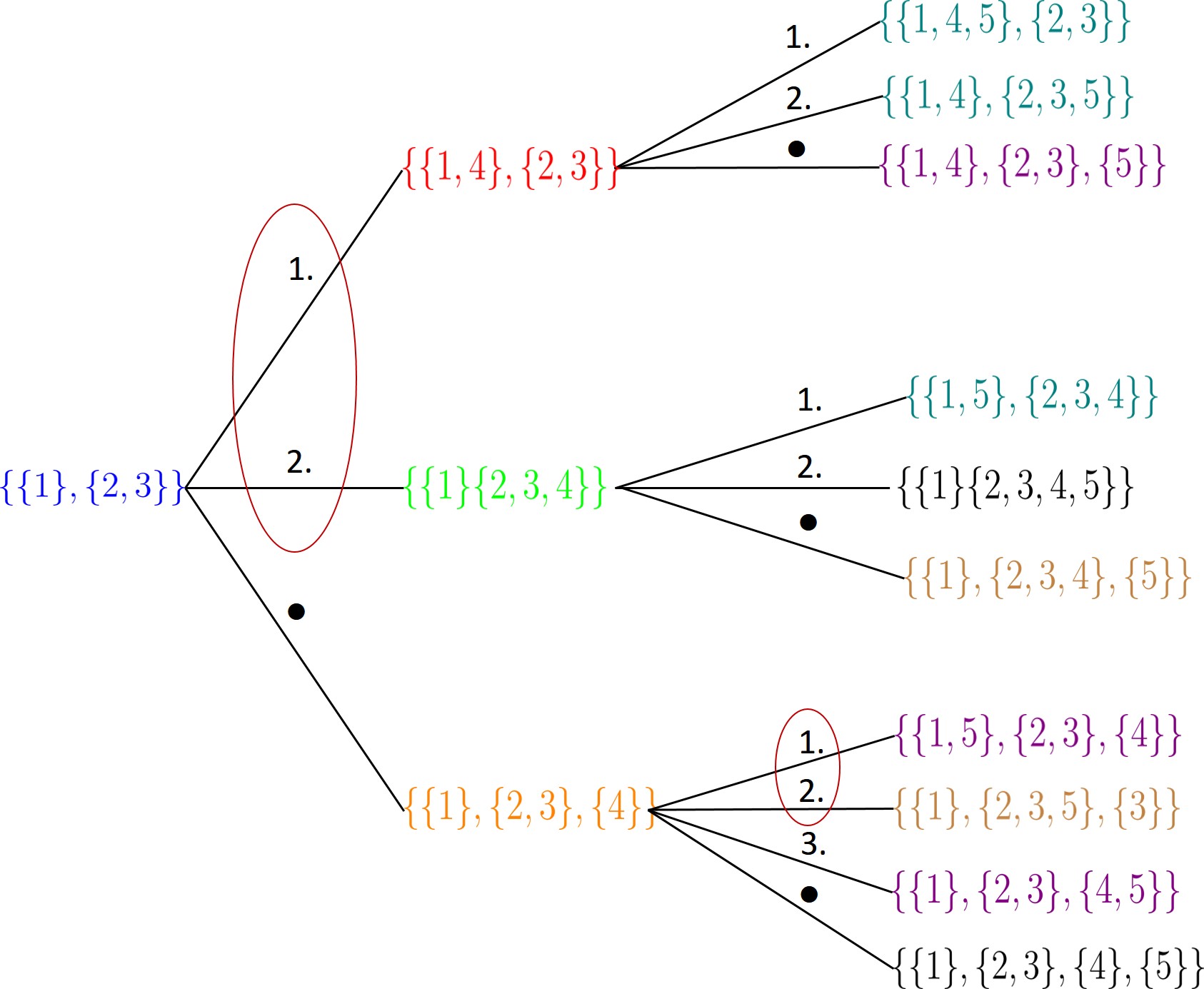}
\end{center}
\end{figure}

Figure~\ref{snapshot1} does suggest though that exchangeability might at least satisfy ``reverse'' Bayesianism. Observe that the ratio of probabilities for objects 1. and 2. remains the same upon sampling novelty ``$\bullet$''. Yet, this turns out to be an artifact of considering only the first couple of periods. Consider instead the snapshot in Figure~\ref{snapshot2} showing the evolution from partition $\{\{1\}, \{2, 3\}\}$. Again, we indicate the object drawn above each edge. Moreover, all partitions of the same color (except black) have the same probability as implied by partition exchangeability. Focus on the relative likelihoods for drawing objects 1. versus 2. at partition \{$\{1\}, \{2, 3\}\}$ and after novelty has been drawn (i.e., ``$\bullet$'') at partition $\{\{1\},\{2, 3\}, \{4\}\}$. For convenience, we circled the relevant draws. Unless further conditions are imposed, partition exchangeability per se does not require that these relatively likelihoods must be the same.

Partition exchangeability satisfies though a very weak version of ``reverse'' Bayesianism. Namely, consider the special sample path along which \emph{in each period} a novelty is drawn. Now, along this path it must be true that the predictive probabilities for known objects associated with partition exchangeable beliefs maintain the same ratio. In fact, at each stage along such special sample path the probabilities over known objects are uniform.

The De Morgan rule, the one-parameter prediction rule associated with the Ewens' sampling formula, and the two-parameter rule of Pitman (1995) and Zabell (1997) are three examples of partition exchangeable belief updating rules that satisfy ``reverse'' Bayesianism, Bayesianism, and ``extended'' Bayesianism. In particular, while in general there can be many ``reverse'' Bayesian updates, the aforementioned prediction rules pin down a ``reverse'' Bayesian update. Theorems~\ref{Zabell_SEU} and~\ref{Ewens_SEU} imply immediately the following:

\begin{corollary} The collection of partition-dependent preferences $\{\succeq_{\bm{\pi}^T}\}_{\bm{\pi}^T \in \Pi^T, T \geq 1}$ associated with the two-parameter model of Theorem~\ref{Zabell_SEU}, the one-parameter model of Theorem~\ref{Ewens_SEU}, and the De Morgan model of Theorem~\ref{Ewens_SEU} with $\theta = 1$ all satisfy ``reverse'' Bayesianism, Bayesianism, and ``extended'' Bayesianism.
\end{corollary}

It follows that together Properties~\ref{frequency_dependence} and~\ref{frequency_dependence_new} imply Properties~\ref{novelty_sampling_consistency} to~\ref{consistency_for_novelty} given Assumption~\ref{assSEU} and Properties~\ref{PE} and~\ref{full_support}. 

Kuipers (1973) proposed a prediction rule for novel objects that, as Zabell (1997, p. 260-261) remarks, is not consistent with partition exchangeable beliefs. Kuipers''s prediction rule states that
\begin{eqnarray*}\mbox{Cond. prob. of a novel object in } T+1 & = & \frac{\mbox{Number of known objects} + \frac{\delta}{2}}{T + \delta}
\end{eqnarray*} and
$$\mbox{Cond. prob. of known object} j \mbox{ in } T+1 =  \frac{T - \mbox{Number of known objects} + \frac{\delta}{2}}{T + \delta}$$
$$\times \frac{\mbox{Number of times object } j\mbox{ occurred} + \frac{\lambda}{\mbox{Number of known objects}}}{T + \frac{\lambda}{\mbox{Number of known objects}}}$$ for $\lambda, \delta \in \mathbb{R}_{++}$.
The ratio of probabilities for distinct known objects $i$ and $j$ is
$$\frac{\mbox{Number of times object } i\mbox{ occurred} + \frac{\lambda}{\mbox{Number of known objects}}}{\mbox{Number of times object } j\mbox{ occurred} + \frac{\lambda}{\mbox{Number of known objects}}}$$ This ratio is increasing in the number of known objects (i.e., upon discoveries of novel objects) if and only if $$\mbox{Number of times object } i\mbox{ occurred} > \mbox{Number of times object } j\mbox{ occurred}.$$ Although the occurrence of novel objects does not preserve the ratio of probabilities for known objects, its change is monotone in the occurrence of novel objects.

It is interesting that Kuipers proposes a prediction rule that does not satisfy ``reverse'' Bayesianism. It is intuitive that a violation of ``reverse'' Bayesianism may be embraced when the decision maker becomes aware of events that also shatter the interpretation of events that the decision maker has been previously aware of like when having transformative experiences of which one was previously unaware of (i.e., Paul, 2014). It can also be violated in strategic settings when being made aware of some types by some opponent allows the player to learn about the types of opponents that she has been previously been aware of. However, these features are absent in the sampling problem. Notable, Kuipers' rule does not satisfy partition exchangeability. It may hint at the fact that Kuipers did not have sampling in ``similar'' circumstances in mind and this lack of symmetry may have also motivated lack of ``reverse'' Bayesianism.

\section{Application to Unawareness Structures\label{ASEU}}

The ``delabeled'' subjective expected utility allowed us to capture partition exchangeability and the two-parameter prediction rule in terms of properties on preferences of the decision maker and thus allowed for a subjective interpretation of the probabilities. However, the approach is somewhat artificial when viewed at the interim basis when some objects have already been sampled. At that point, the decision maker is able to reason about all previously sampled objects in addition to the occurrence of novelty. That's is, to model the subjective perspectives of the decision maker, it would be conceptually desirable to have a hybrid approach that starts at a completely ``delabeled'' description and then adds labels steps-by-step as the process evolves. We can use unawareness structures for this purpose (Heifetz, Meier, and Schipper, 2006, 2008, 2013, Schipper, 2013). The primitive of unawareness structures is a lattice of spaces ordered by their richness together with projections from richer to poorer spaces. When focusing on the predictive probabilities, in our context it is natural to think of each spaces as a distinct subset of objects together with novelty, ``$\bullet$''.
\begin{figure}[h!]
\caption{Example of an Unawareness Structure with Predictive Probabilities Implied by the Two-Parameter Model\label{unawareness_structure}}
\begin{center}
\includegraphics[scale=0.4]{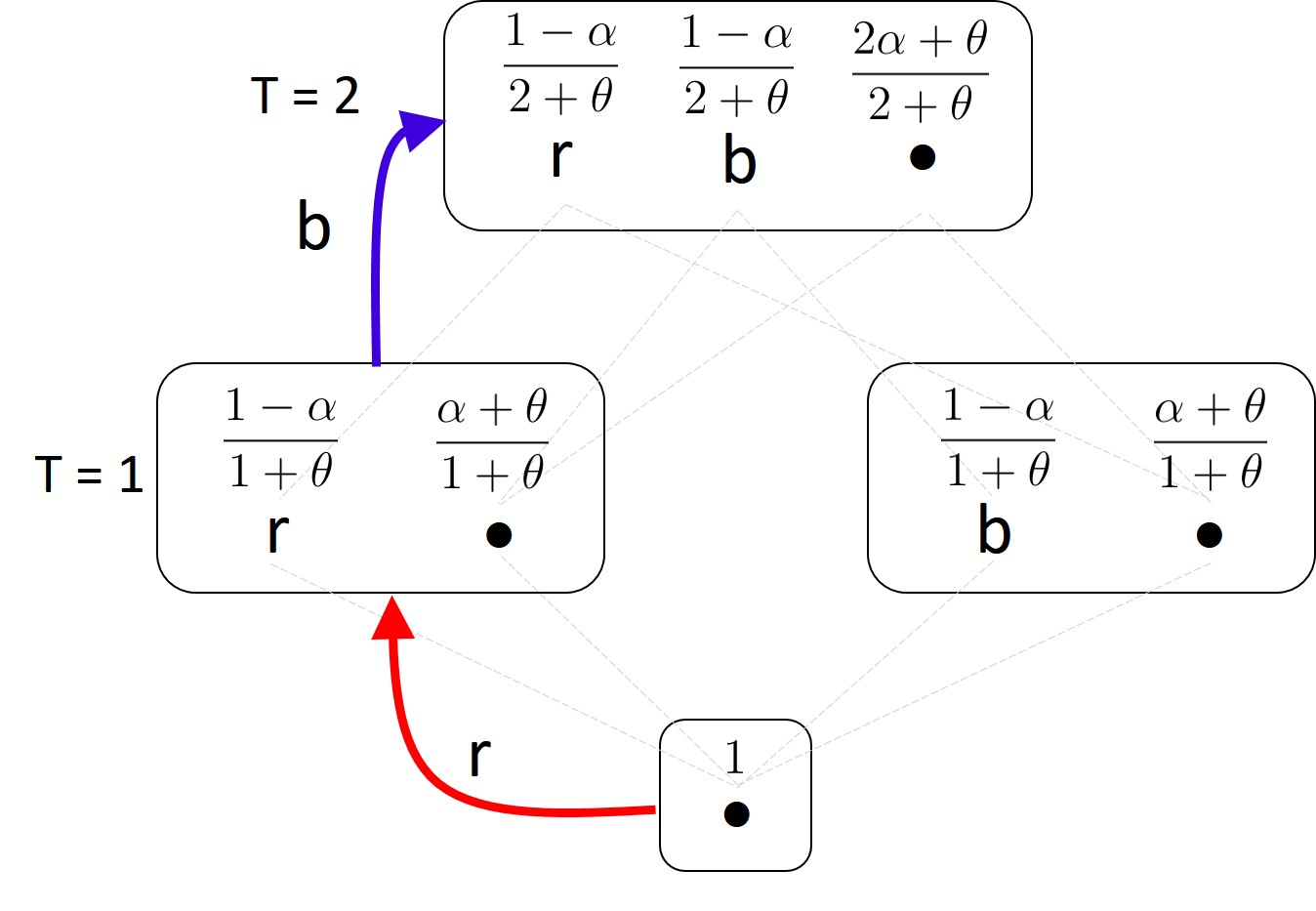}
\end{center}
\end{figure}

More formally, for any $T$ and sequence of drawn objects $\bm{s}^T = (s_1, ..., s_T)$, let the associated subjective state space be $S(\bm{s}^T) = \{s_t\}_{t = 1}^T \cup \{\bullet\}$, where $\{s_t\}_{t = 1}^T$ denotes the set of objects drawn till period $T$. These are the only objects in addition to novelty that a decision maker can form beliefs over conditional on $\bm{s}^T$. Clearly, when considering sequences of sampled objects, the possible subjective state spaces that may arise are partially ordered by set inclusion. In fact, they form a complete lattice. For any two spaces $S$ and $S'$ with $S \subseteq S'$, the projection is defined such that all objects in $S'$ that are also contained in $S$ are mapped one-to-one and any other objects contained in $S' \setminus S$ are mapped to $\bullet$. Also, $\bullet$ is mapped to $\bullet$.

As a simple example, consider the lattice of four spaces in Figure~\ref{unawareness_structure}. Projections are indicated by gray dashed lines. The smallest space contains only $\bullet$. This is the initial space at $T = 0$. Suppose there is a red ball $r$ and a blue ball $b$ of which the decision maker is initially unaware. Depending on which ball is drawn at $T = 1$, the state space after $T=1$ is either the left or the right space given by $\{r, \bullet\}$ and $\{b, \bullet\}$, respectively. Suppose that $r$ is drawn as indicated by the red arrow. Then the predictive probabilities after $T = 1$ of the two-parameter model are as written above the states in the left space. If $b$ is drawn in $T = 2$ as indicated by the blue arrow, then the predictive probabilities of the two-parameter model are as written above the states in the upmost space. Ex post (e.g., after $T = 2$) the decision maker may also be able to reason about what would have been her predictive probabilities at the end of $T = 1$ if initially the blue ball had been drawn.  These counterfactual considerations may not be relevant in a single-person decision problem but may become important for strategic reasoning in game theoretic models. In any case, the example demonstrates how the models discussed in this paper can give rise to subjective probabilities in unawareness structures when the lattice of spaces is interpreted as arising from a dynamic process of sampling objects.

\section{Discussion\label{discussion}}

Sampling is naturally a domain in which frequencies and objective stochastic processes play an obvious role. Yet, as the discussion of de Finetti's Theorem by Kreps (1988, pp. 145) illustrates, it is also a domain of subjective beliefs. What does it mean to an analyst that she believes that partitions of sampling times are exchangeable? What does it mean that she beliefs a novelty occurs in the next period? Choices can reveal such beliefs even if we cannot anticipate particular novelties, their characteristics etc. And such beliefs may be entertained when the underlying probability process is unknown to the analyst and she is a subjective expected utility maximizer.

De Finetti famously questioned the doxiastic existence of objective probabilities. The central result in the theory of exchangeable random partitions is Kingman's characterization as a mixture of paint-ball processes. That is, beliefs over infinite exchangeable random partitions can be captured by a mixture of distributions that have atomistic parts, corresponding to objects that occur frequently, and a continuous part, corresponding to objects that occur only once. This is a version of de Finetti's Theorem (see Aldous, 1985, for a proof essentially using de Finetti's Theorem). Like de Finetti's theorem, Kingman's Theorem can be viewed as infinite exchangeable random partitions giving rise to a subjective belief, the mixture over what is called paint-ball processes. In this paper, we are less interested in limit beliefs as the sampling periods go to infinity. More relevant to decision making, we consider beliefs at any finite sampling period although limit beliefs can be easily stated for the models we consider; see for instance, Pitman (1995) for the two-parameter model. Nevertheless, Kingman's Theorem emphasizes the a key assumption behind our approach: When we delabel objects in the world of random partitions, the decision maker when sampling novelties only faces uncertainty about the distribution but his beliefs are not shattered by the novelties he discovers because the beliefs in this model are not about the novelties and their characteristics or attributes themselves but just about the occurrence of novelty, something that he is able to form beliefs about even ex ante.

There might be extensions that allow for a closer connection between objective innovation processes and subjective beliefs. Similar to De Castro and Al-Najjar (2014) for the case of exchangeable sequences, it should be possible to consider parametric representations of preferences in our setting. Or it might be possible to incorporate objective information into subjective expected utility analogous to the approach by Cerreia-Vioglio et al. (2013). We leave these considerations to future work.

In situations under awareness of unawareness, the degree of uncertainty is rather substantial. Thus, it might make sense to consider ambiguous subjective beliefs rather than subjective probability measures. Walley (1996) studies sampling of novel objects in a statistical approach involving imprecise probabilities. Although he does not consider partitions of sampling times, he argues that in the face of awareness of unawareness, the analysis cannot depend on the sample space. Perhaps an ``ambiguous'' version of Kingman's Theorem can be proved analogous to the ambiguous version of de Finetti's theorem by Epstein and Seo (2010). Again, this must be left to future work.

We were made aware of the discovery of species problem by the discussion of novelty by Witt (2009). He criticizes Zabell (1992) as just pertaining to novelties of minimal degree. According to Witt (2009), ``the ex ante degree of novelty is minimal ... if the generative operation is known and the interpretative operation is trivial in the sense that the meaning of all its possible outcomes is known beforehand.'' Generative operation means the process of producing novelty. Interpretative operation refers to the integration of the phenomenon into emerging or existing concepts. Witt (2009) claims that in Zabell's work ``all possible outcomes of the generative operation can be anticipated''. This is clearly not the case. In fact, by delabeling outcomes and just considering partitions of sampling times instead, the approach avoids specifying beforehand particular outcomes. In other words, in the approach there is no attempt to specify the sample space or its outcomes before. What can be anticipated is the event that either a previously sampled outcome is re-sampled or a novel outcome is sampled. But the space of partitions of sampling times should not be confused with the underlying sample space of objects. What is implicitly known in the discovery of species problem though is the objective stochastic process of sampling novelty. In our approach, this is replaced by subjective beliefs. Witt (2009) further notes that ``the interpretative operation is trivial.'' But this is not due to knowledge of all possible outcomes beforehand. Rather by delabeling outcomes, the approach gives up entirely on the interpretative operation and simply remains silent on it. The approach is not a theory of novelties but rather a theory of the occurrences of novelty.

Using the discovery of species problem as a model of sampling novelties can be criticized though from different angles. The underlying assumption is partition exchangeability. It models the assumption that novelties are sampled in ``similar'' circumstances. This is clearly a matter of perspective. Every successful inventor will rightfully claim the special circumstances of her unique invention or discovery. Moreover, there could be discoveries of radical novelties after which the world is not the same anymore. Yet, from the perspective of a grant making agency or administration of academic research, institutes, labs, researchers etc. are treated ``similar''. And even an inventor will have to admit that there is a daily routine in day-to-day research activities that occasionally lead to discoveries. Thus from such perspectives, the assumption of exchangeable partitions may be justified.

Another critique is that the model forces the decision maker to be aware that she is unaware. The model does not allow for initial unawareness of unawareness and awareness of unawareness only upon a first occurrence of novelty. The decision maker anticipates from the very beginning the occurrence novelty. On top of it, she must believe from the very beginning that in the limit, novelty will have occurred an infinite number of times. Such a state of mind is a rather special case of awareness of unawareness. For general epistemic models of awareness of unawareness, see Schipper (2024), Halpern and Rego (2013), Board and Chung (2021), and Sillari (2008).

\appendix

\section{Proofs\label{proofs}}

\subsection*{Proof of Proposition~\ref{PE_result}}

By Assumption~\ref{assSEU} for any $T > 1$, $\bm{\pi}^T \in \Pi^T$, we have for some $z_{\bm{\pi}^T} \in \mathbb{R}$ that
$$1_{[\bm{\pi}^T]} 0 \sim z_{\bm{\pi}^T}$$ if and only if
 \begin{eqnarray*} \int_{\Omega} u(1_{[\bm{\pi}^T]} 0) d \mu & = & \int_{\Omega} u(z_{\bm{\pi}^T}) d \mu \\
\int_{[\bm{\pi}^T]} u(1) d \mu + \int_{\Omega \setminus [\bm{\pi}^T]} u(0) d \mu & = & u(z_{\bm{\pi}^T}) \mu(\Omega) \\
u(1) \mu([\bm{\pi}^T]) + u(0) \mu(\Omega \setminus [\bm{\pi}^T]) & = & u(z_{\bm{\pi}^T}) \\
\mu([\bm{\pi}^T]) & = & \frac{u(z_{\bm{\pi}^T})}{u(1)}
\end{eqnarray*} where the last line uses the fact that by Assumption~\ref{assSEU} we can normalize $u$ such that $u(0) = 0$ and we have by $u$ being strictly increasing on $\mathbb{R}$ that $u(1) \neq 0$.

By similar arguments, for $\bm{\tilde{\pi}}^T \in \Pi^T$,
\begin{eqnarray*} \mu([\bm{\tilde{\pi}}^T]) & = & \frac{u(z_{\bm{\tilde{\pi}}^T})}{u(1)}
\end{eqnarray*}

For ``$\Rightarrow$'', assume Property~\ref{PE} and let $a^T(\bm{\pi}^T) = a^T(\bm{\tilde{\pi}}^T)$. Observe that above arguments imply $u(z_{\bm{\pi}^T}) = u(z_{\bm{\tilde{\pi}}^T})$. The conclusion now follows. For ``$\Leftarrow$'', assume $a^T(\bm{\pi}^T) = a^T(\bm{\tilde{\pi}}^T)$ and observe that above arguments imply Property~\ref{PE}.

This proves Proposition~\ref{PE_result} (i). For (ii), apply analogous arguments to the partition conditional preferences. \hfill $\Box$

\subsection*{Proof of Theorem~\ref{Zabell_SEU}}

We start with some preliminaries. For any $T \geq 1$, let $p^T \in \Delta(\Pi^T)$ be a probability measure on $\Pi^T$. For any $T, T' \geq 1$ with $T' \leq T$, define the surjective projection $\rho^T_{T'}: \Pi^T \longrightarrow \Pi^{T'}$ by $\rho^T_{T'}(\bm{\pi}^T) = \bm{\pi}^T_{|T'}$ and $\rho^T_T$ the identity on $\Pi^T$. A collection of probability measures $(p^T)_{T = 1, ...}$ is consistent if $p^T \circ (\rho^T_{T'})^{-1} = p^{T'}$ for any $T, T' \geq 1$ with $T' \leq T$.

Let $\Pi^{\infty}$ the space of all infinite partitions $\bm{\pi}^{\infty}$ of $\mathbb{N}_+$  generated by consistent collections of partitions $(\bm{\pi}^T)_{T \geq 1}$ with $\bm{\pi}^{T}_{\mid T'} = \bm{\pi}^{T'}$ for all $T > T' \geq 1$. By the Kolmogorov extension theorem (Aliprantis and Border, 2006, Chapter 15.6), there exists a unique countable additive probability measure $p^{\infty} \in \Delta(\Pi^{\infty})$ with marginals $p^{T}$ for any $T \geq 1$.

We say that $(p^T)_{T \geq 1}$ is partition exchangeable if for any $T \geq 1$, $a^T(\bm{\pi}^T) = a^T(\bm{\tilde{\pi}}^T)$ implies $p^T(\bm{\pi}^T) = p^T(\bm{\tilde{\pi}}^T)$. We say that $p^{\infty}$ is infinite partition exchangeable if it is the extension of partition exchangeable consistent probability measures $(p^T)_{T \geq 1}$.

Consider $p^{\infty}$ with marginals $(p^T)_{T \geq 1}$ and the following conditions:
\begin{itemize}
\item[(A)] $p^{\infty}$ infinite partition exchangeable.
\item[(B)] For $T \geq 1$, $\bm{\pi}^T \in \Pi^T$,
\begin{eqnarray}
p^T(\bm{\pi}^T) & > & 0. \label{c1}
\end{eqnarray}
\item[(C)] For $T \geq \ell \geq 1$, $\bm{\pi}^T = \{\bm{\pi}^T_1, ..., \bm{\pi}^T_{\ell}\} \in \Pi^T$, and $j = 1, ..., \ell$,
\begin{eqnarray} p^{T+1}\left(\bm{\pi}^{T+1}_j = \bm{\pi}^T_j \cup \{T+1\} \mid \bm{\pi}^T \right)  & = &  f(|\bm{\pi}^T_j|, T). \label{c2}
\end{eqnarray}
\item[(D)] For $T \geq 1$ and $\bm{\pi}^T \in \Pi^T$,
\begin{eqnarray}
p^{T+1}\left(\bm{\pi}^{T+1}(T+1) = \{T+1\} \mid \bm{\pi}^T \right) & =  & g(|\bm{\pi}^T|, T) \label{c3}.
\end{eqnarray}
\end{itemize}

\begin{lemma}[Zabell, 1997]\label{Zabell}  Conditions (A) to (D) imply that there exist $\alpha$ and $\theta$ with $\alpha \in [0, 1)$ and $\theta > - \alpha$ such that
\begin{eqnarray} f(|\bm{\pi}^T_j|, T) & = & \frac{|\bm{\pi}^T_j| - \alpha}{T + \theta} \\
g(|\bm{\pi}^T|, T) & = & \frac{\alpha |\bm{\pi}^T| + \theta}{T + \theta}.
\end{eqnarray}
\end{lemma}

We continue with the proof of Theorem~\ref{Zabell_SEU}. First, we consider direction ``$\Rightarrow$''. We verify conditions (A) to (D). 

Condition (A): By Assumption~\ref{assSEU}, $\succeq$ has a countable additive subjective expected utility representation. Denote by $\mu$ the associated probability measure. For any $T$ and $\bm{\pi}^T \in \Pi^T$, let $p^T(\bm{\pi}^T) := \mu([\bm{\pi}^T])$. This defines a probability in $\Delta(\Pi^T)$. We claim that for all $T$ and $T' \leq T$, $\bm{\pi}^{T'} \in \Pi^{T'}$, 
$$\mu\left(\left[(\rho^T_{T'})^{-1}(\bm{\pi}^{T'})\right]\right) = \mu\left(\left[\bm{\pi}^{T'}\right]\right).$$ Note that $\left[(\rho^T_{T'})^{-1}(\bm{\pi}^{T'})\right]$ is measurable because it is a finite union of measurable events, i.e., $\left[(\rho^T_{T'})^{-1}(\bm{\pi}^{T'})\right] = \bigcup_{\bm{\pi}^T \in (\rho^T_{T'})^{-1}(\bm{\pi}^{T'})} \left[\bm{\pi}^{T}\right]$. 

We will show $\left[(\rho^T_{T'})^{-1}(\bm{\pi}^{T'})\right] = \left[\bm{\pi}^{T'}\right]$. 

Consider the case ``$\supseteq$'': For any $\omega \in \left[\bm{\pi}^{T'}\right]$, $\pi^{T'}(\omega) = \bm{\pi}^{T'}$. By the partition functions $(\pi^T)_{T = 1, 2, ...}$ satisfy consistency, we have $\pi^T(\omega)_{|T'} = \pi^{T'}(\omega)$. Since $\pi^T$ is a function with codomain $\Pi^T$, there exist $\bm{\pi}^T \in \Pi^T$ such that $\pi^T(\omega) = \bm{\pi}^T$. This is the case if and only if $\omega \in [\bm{\pi}^T]$. Consistency of $(\pi^T)_{T = 1, 2, ...}$ implies that $\bm{\pi}^T \in (\rho^T_{T'})^{-1}(\bm{\pi}^{T'})$. Thus, $\omega \in \left[(\rho^T_{T'})^{-1}(\bm{\pi}^{T'})\right]$. 

Next consider the case ``$\subseteq$'': Let $\omega \in \left[(\rho^T_{T'})^{-1}(\bm{\pi}^{T'})\right]$ implies $\omega \in \left[\bm{\pi}^{T}\right]$ for some $\bm{\pi}^T \in (\rho^T_{T'})^{-1}(\bm{\pi}^{T'})$. Observe that $\omega \in \left[\bm{\pi}^{T}\right]$ if and only if $\pi^T(\omega) = \bm{\pi}^T$. Since partition functions $(\pi^T)_{T = 1, 2, ...}$ satisfy consistency, $\pi^T(\omega)_{|T'} = \pi^{T'}(\omega)$. Since $\bm{\pi}^T \in (\rho^T_{T'})^{-1}(\bm{\pi}^{T'})$, $\pi^{T'}(\omega) = \bm{\pi}^{T'}$. Thus, $\omega \in \left[\bm{\pi}^{T'}\right]$. 

It follows now that the collection of probability measures $(p^T)_{T \geq 1}$ is consistent. By Property~\ref{PE} and Proposition~\ref{PE_result}, for $T \geq 1$, $p^T$ is partition exchangeable. By above arguments, there exists a unique extension $p^{\infty}$ that is infinite partition exchangeable. This verifies Condition (A).

Condition (B): Using Assumption~\ref{assSEU}, Property~\ref{full_support} implies Condition (B).

Condition (C): By Assumption~\ref{assSEU}, in the following we can conveniently normalize $u$ w.l.o.g. such that $u(0) = 0$. 

By Assumption~\ref{assSEU}, arguments analogous to the proof of Proposition~\ref{PE_result} imply by Property~\ref{frequency_dependence} for all $T \geq 1$, $k \in \{1, ...T\}$, $\bm{\pi}^T, \bm{\tilde{\pi}}^T \in \Pi^{T, k}$,
$$\mu\left(\left[|\pi^{T+1}(\cdot)(T+1)| = k + 1\right] \mid [\bm{\pi}^T] \right) =  \mu\left(\left[|\pi^{T+1}(\cdot)(T+1)| = k + 1\right] \mid [\bm{\tilde{\pi}}^T] \right).$$
Note that since $\bm{\pi}^T \in \Pi^{T, k}$, it must be that $\left[|\pi^{T+1}(\cdot)(T+1)| = k + 1\right] \cap [\bm{\pi}^{T}] = \\ \left[|\pi^{T+1}_j(\cdot)| = |\pi^{T}_j(\cdot)| + 1\right] \cap [\bm{\pi}^{T}] = \left[\pi^{T+1}_j(\cdot) = \pi^T_j(\cdot) \cup \{T+1\}\right] \cap \left[\bm{\pi}^T\right]$ for some $j \leq T$ with $|\bm{\pi}^T_j | = k$. Since this holds for any partition $\bm{\pi}^T \in \Pi^{T, k}$, we must have that $$\mu\left(\left[|\pi^{T+1}_j(\cdot)| = |\pi^{T}_j(\cdot)| + 1\right] \mid [\bm{\pi}^T]\right) = f(k, T)$$ for some function $f$ of $k$ and $T$. This verifies Condition (C).

For Condition (D), note that by Assumption~\ref{assSEU} and arguments analogous to the proof of Proposition~\ref{PE_result} we have by Property~\ref{frequency_dependence_new} that for $T \geq 1$ and $\bm{\pi}^T, \bm{\tilde{\pi}}^T \in \Pi^{T}$ with $|\bm{\pi}^T| = |\bm{\tilde{\pi}}^T|$,
$$\mu\left(\left[|\pi^{T+1}(\cdot)(T+1)| = 1\right] \mid [\bm{\pi}^T] \right) =  \mu\left(\left[|\pi^{T+1}(\cdot)(T+1)| = 1\right] \mid [\bm{\tilde{\pi}}^T] \right).$$
Observe that $\left[|\pi^{T+1}(\cdot)(T+1)| = 1 \right] \cap \left[\bm{\pi}^T\right] = \left[|\pi^{T+1}(\cdot)| = |\pi^T(\cdot)| + 1 \right] \cap \left[\bm{\pi}^T\right] \\ = \left[\pi^{T+1}(\cdot)(T+1) = \{T+1\}\right] \cap \left[\bm{\pi}^T\right]$. Since this holds for any  partition $\bm{\pi}^T, \bm{\tilde{\pi}}^T \in \Pi^{T}$ with $|\bm{\pi}^T| = |\bm{\tilde{\pi}}^T|$, it must be that $$\mu\left(\left[|\pi^{T+1}(\cdot)| = |\pi^T(\cdot)| + 1 \right] \mid \left[\bm{\pi}^T\right]\right) = g(|\bm{\pi}^T|, T)$$ for some function $g$ of the cardinality of partition $\bm{\pi}^T$ and $T$. This verifies Condition (D).

It now follows from Lemma~\ref{Zabell} that the partition-conditional subjective expected utility representation follows the prediction rule of Theorem~\ref{Zabell_SEU}.

Now consider direction ``$\Leftarrow$''. For any $T \geq 1$ and $\bm{\pi}^T \in \Pi^T$, define $\mu( \cdot \mid [\bm{\pi}^T]) := p^{T+1}(\cdot \mid \bm{\pi}^T)$. By Assumption~\ref{assSEU} there exist a collection of partition conditional preferences $\{\succeq_{\bm{\pi}^T}\}_{\bm{\pi}^T \in \Pi^T, T \geq 1}$.

Pitman (1995, Proposition 9) shows that the prediction rule implies equation~(\ref{EPF_two_parameters}). Assumption~\ref{assSEU} implies now Property~\ref{PE}. 

Property~\ref{full_support} is implied from the range of parameters of the prediction rule of Theorem~\ref{Zabell_SEU} and Assumption~\ref{assSEU}. In particular, conditional on any partition, any previously encountered object has strict positive probability. A novel object has strict positive probability as well. Thus, ex ante no partition is ruled.

Reversing the arguments in the proof of Conditions (C) and (D) of the part ``$\Rightarrow$'' above proves the necessity of Properties~\ref{frequency_dependence} and~\ref{frequency_dependence_new}, respectively. 

Finally, we show how to calibrate the parameters of the prediction rule from choices. Let $z \in \mathbb{R}$ be defined by the choice behavior $1_{\left[\{\{1\}, \{2, 3\}, \{4\}\}\right]} 0 \sim_{\{\{1\}, \{2, 3\}\}} z$ and $k \in \mathbb{R}$ be defined by choice behavior $1_{\left[\{\{1, 4\}, \{2, 3\}\}\right]} 0 \sim_{\{\{1\}, \{2, 3\}\}} k$. We continue to assume that $u$ is normalized such that $u(0) = 0$. By arguments similar to the proof of Proposition~\ref{PE_result} we obtain
\begin{eqnarray*}
\mu(\{\{1\}, \{2, 3\}, \{4\}\} \mid \{\{1\}, \{2, 3\}\}) & = & \frac{u(z)}{u(1)}
\end{eqnarray*} and
\begin{eqnarray*} \mu(\{\{1, 4\}, \{2, 3\}\} \mid \{\{1\}, \{2, 3\}\}) & = & \frac{u(k)}{u(1)}
\end{eqnarray*} Using the prediction rules of equations~(\ref{known}) and~(\ref{novel}), respectively, we obtain
\begin{eqnarray} \frac{u(k)}{u(1)} & = & \frac{1 - \alpha}{3 + \theta}
\\
\frac{u(z)}{u(1)} & = & \frac{2 \alpha + \theta}{3 + \theta}
\end{eqnarray} Solving these two equations for $\alpha$ and $\theta$ yields the expressions in equations~(\ref{alpha}) and~(\ref{theta}), respectively.\hfill $\Box$

\subsection*{Proof of Proposition~\ref{risk_neutrality}}

We prove just the case for the two-parameter model. The other cases are direct corollaries. Since $\succeq_{\bm{\pi}^T}$ has a subjective expected utility representation, choice $$(T+\theta)_{[\bullet]} 0 \sim_{\bm{\pi}^T} (\alpha |\bm{\pi}^T| + \theta)$$ is equivalent to
$$\int_{\Omega} u\left((T+\theta)_{[\bullet]} 0\right) d \mu(\cdot \mid \bm{\pi}^T) = \int_{\Omega} u\left((\alpha |\bm{\pi}^T| + \theta)\right) d \mu(\cdot \mid \bm{\pi}^T).$$ Under risk neutrality we can chose $u(x) = x$. Thus, we obtain
\begin{eqnarray*} \int_{\Omega} (T+\theta)_{[\bullet]} 0 d \mu(\cdot \mid \bm{\pi}^T) & = & \int_{\Omega} (\alpha |\bm{\pi}^T| + \theta) d \mu(\cdot \mid \bm{\pi}^T)\\
(T + \theta) \mu([\bullet] \mid \bm{\pi}^T) & = & (\alpha |\bm{\pi}^T| + \theta) \\
\mu([\bullet] \mid \bm{\pi}^T) & = & \frac{\alpha |\bm{\pi}^T| + \theta}{T+\theta}.
\end{eqnarray*} The case $$(T+\theta)_{[j]} 0 \sim_{\bm{\pi}^T} (|\bm{\pi}_j^T| - \alpha)$$ follows by analogous arguments. \hfill $\Box$

\subsection*{Proof of Proposition~\ref{reverse_Bayesianism}}

We prove part (i): Let $\left\{\succeq_{\bm{\pi}^T}\right\}_{\bm{\pi}^T \in \Pi^T, T \geq 1}$ satisfy Assumption~\ref{assSEU}. Recall Property~\ref{novelty_sampling_consistency}: for any $T$ and $\bm{\pi}^T \in \Pi^T$ and $i, j \leq |\bm{\pi}^T|$, there are $x, y \in \mathbb{R}$ such that
\begin{eqnarray*} x_{\left[T+1 \in \pi^{T+1}_i(\cdot)\right]} 0 \sim_{\bm{\pi}^T} y_{\left[T+1 \in \pi^{T+1}_j(\cdot)\right]}0 & \mbox{ iff } & x_{\left[T+2 \in \pi^{T+2}_i(\cdot)\right]} 0 \sim_{\bm{\pi}^{T+1}} y_{\left[T+2 \in \pi^{T+2}_j(\cdot)\right]}0
\end{eqnarray*} for $\bm{\pi}^{T+1} = \bm{\pi}^T \cup \{T+1\}$. By Assumption~\ref{assSEU}, the left-hand side is equivalent to
\begin{eqnarray*} \int_{\Omega}  u\left(x_{\left[T+1 \in \pi^{T+1}_i(\cdot)\right]} 0\right) d \mu\left(\cdot \mid \left[\bm{\pi}^T\right]\right) & = & \int_{\Omega}  u\left(y_{\left[T+1 \in \pi^{T+1}_j(\cdot)\right]} 0\right) d \mu\left(\cdot \mid \left[\bm{\pi}^T\right]\right) \\
u(x) \mu\left(\left[T+1 \in \pi^{T+1}_i(\cdot)\right] \mid \left[\bm{\pi}^T\right]\right)  & = & u(y) \mu\left(\left[T+1 \in \pi^{T+1}_j(\cdot)\right] \mid \left[\bm{\pi}^T\right]\right)
\end{eqnarray*} By Property~\ref{full_support} and Assumption~\ref{assSEU}, probabilities must be nonzero. Thus, we can rewrite the equation as
\begin{eqnarray*}
\frac{\mu\left(\left[T+1 \in \pi^{T+1}_i(\cdot)\right] \mid \left[\bm{\pi}^T\right]\right)}{\mu\left(\left[T+1 \in \pi^{T+1}_j(\cdot)\right] \mid \left[\bm{\pi}^T\right]\right)} & = & \frac{u(y)}{u(x)}
\end{eqnarray*} By analogous arguments, the r.h.s. is equivalent to
\begin{eqnarray*}
\frac{\mu\left(\left[T+2 \in \pi^{T+2}_i(\cdot)\right]\mid \left[\bm{\pi}^{T+1}\right]\right)}{\mu\left(\left[T+2 \in \pi^{T+2}_j(\cdot)\right]\mid \left[\bm{\pi}^{T+1}\right]\right)} & = & \frac{u(y)}{u(x)}
\end{eqnarray*} ``Reverse'' Bayesianism now follows.

Parts (ii) and (iii) follow analogously by replacing Property~\ref{novelty_sampling_consistency} in above arguments with Property~\ref{sampling_consistency} or \ref{consistency_for_novelty}, respectively. \hfill $\Box$

\section{SEU Representation\label{Wakker}}

We the sake of completeness, we state the properties and theorem characterizing Assumption~\ref{assSEU}. We find it convenient to make use of Wakker (1989, Chapter V.5). Let $C$ be a nonempty connected separable topological space of outcomes or consequences endowed with the Borel $\sigma$-algebra $\Sigma_C$. For any outcomes $x, y \in C$, we let $x \curlyvee y = x$ if $x \succeq y$ and $x \curlyvee y = y$ otherwise. Similarly, $x \curlywedge y = x$ if $x \preceq y$ and $x \curlywedge y = y$ otherwise. For any act $f$ and consequence $x$, the \emph{above truncation} is the act $f \curlyvee x$ defined by $(f \curlyvee x)(\omega) = f(\omega) \curlyvee x$ for all $\omega \in \Omega$. Define analogously the \emph{below truncation}. A set of acts is truncation closed w.r.t. $\succeq$ if for every act in that set and every consequence also the above truncation and below truncation is contained in that set. We say that an act $f$ is \emph{bounded} w.r.t. $\succeq$ if there exist consequences $x, y \in C$ such that $x \succeq f \succeq y$.

We say that $\langle \langle \Omega, \Sigma_{\Omega}\rangle, \langle C, \Sigma_C \rangle, \mathcal{F}, \succeq \rangle$ is an \emph{abstract choice frame} if $\mathcal{F}$ is the set of all measurable acts $f: \Omega \longrightarrow C$ that include the simple acts, are truncation closed w.r.t. $\succeq$, and are bounded w.r.t. $\succeq$. In the following we fix an abstract choice frame.

\begin{property}[Weak order]\label{weak_order} $\succeq$ is a weak order, i.e., complete and transitive.
\end{property}

\begin{property}[Pointwise monotonicity]\label{pointwise_monotonicity} $\succeq$ is pointwise monotone, i.e., for any $f, g \in \mathcal{F}$, $f(\omega) \succeq g(\omega)$ for all $\omega \in \Omega$ implies $f \succeq g$.
\end{property}

For event $E \in \Sigma_{\Omega}$ and consequences $x, y, z, w \in C$, we write $$xy \unrhd^E zw$$ if there exist simple acts $f, g \in \mathcal{F}$ such that $x_E f \succeq y_E g$ and $z_E f \preceq w_E g$. We write $$xy \rhd^E zw$$ if there exist simple acts $f, g \in \mathcal{F}$ such that $x_E f \succeq y_E g$ and $z_E f \not\succeq w_E g$.

\begin{property}[Event Tradeoff Consistency]\label{event_tradeoff_consistency} There do not exist simple $\succeq$-non-null events $E, F \in \Sigma_{\Omega}$ and consequences $x, y, z, w \in C$ such that $xy \unrhd^E zw$ and $zw \rhd^F xy$.
\end{property}

For any partition $\{E_i\}_{i = 1}^m$ of $\Omega$ and consequences $x^1, ..., x^m \in C$, denote by $x^1_{E_1} x^2_{E_2} ... x^m_{E_m}$ the composite act defined by $x^1_{E_1} x^2_{E_2} ... x^m_{E_m} (\omega) = x^{\ell}$ if $\omega \in E_\ell$.

\begin{property}[Simple-continuity]\label{simple-continuity} $\succeq$ is simple-continuous if for any partition $\{E_i\}_{i = 1}^m$ of $\Omega$ and simple act $f$ measurable w.r.t. $\{E_i\}_{i = 1}^m$, the better set $\{(x^1, ..., x^m) \in C^m : x^1_{E_1} x^2_{E_2} ... x^m_{E_m} \succeq f\}$ and worse set $\{(x^1, ..., x^m) \in C^m : x^1_{E_1} x^2_{E_2} ... x^m_{E_m} \preceq f\}$ are close w.r.t. the product topology on $C^m$.
\end{property}

\begin{property}[Constant-continuity]\label{constant-continuity} $\succeq$ is constant-continuous if for all $f \in F$, the better set $\{ x \in C: x \succeq f\}$ and the worse set $\{x \in C : x \preceq f\}$ are closed.
\end{property}

\begin{property}[Truncation-continuity]\label{truncation-continuity} $\succeq$ is truncation-continuous if for every $f, g \in \mathcal{F}$ with $f \succ g$, there exists $x, y \in C$ such that $f \curlywedge x \succ g$ and $f \succ g \curlyvee y$.
\end{property}

A set of acts is uniformly strongly bounded if there exist $x, y \in C$ such that $x \succeq f(\omega) \succeq y$ for all $\omega \in \Omega$ and acts $f$ in the set.

\begin{property}[Bounded-strict-continuity]\label{boundedly-strict-continuity} $\succeq$ is boundedly strictly continuous if for any uniformly bounded sequence of acts $(f^i)_{i = 1}^\infty$ in $\mathcal{F}$ and any pair of acts $g, h \in \mathcal{F}$ for which $f^i \succeq h$ (resp., $f^i \preceq h$) for all $i$ and $\lim_{i \rightarrow \infty} f^i(\omega) = g(\omega)$ for all $\omega \in \Omega$, we have $g \succeq h$ (resp., $g \preceq h$).
\end{property}

\begin{definition}[Countable-Additive Subjective Expected Utility]\label{SEU} The preference relation $\succeq$ admits a countable-additive Subjective Expected Utility representation if for all $f, g \in \mathcal{F}$, $$f \succeq g$$ if and only if
\begin{eqnarray*} \int_{\Omega} u \circ f \ d \mu & \geq & \int_{\Omega} u \circ g \ d \mu.
\end{eqnarray*}
\normalsize for a continuous utility function $u: C \longrightarrow \mathbb{R}$ and a countable-additive probability measure $\mu \in \Delta(\Omega)$.

If $\Omega$ is simple $\succeq$-null, then $\mu$ is arbitrary and $u$ is constant.

If $\Omega$ is simple $\succeq$-non-null but no two disjoint simple $\succeq$-non-null events exist in $\Sigma_{\Omega}$, then $\mu$ assigns probability $1$ to every simple $\succeq$-non-null event and probability $0$ to every simple $\succeq$-null event, and $u$ is unique up to continuous strictly increasing transformations.

Otherwise, the utility function is unique up to positive affine transformations and $\mu$ is uniquely determined.
\end{definition}

\begin{theorem}[Wakker, 1989]\label{SEU_theorem} Properties~\ref{weak_order} to~\ref{boundedly-strict-continuity} hold if and only if $\succeq$ admits a countable additive subjective expected representation.
\end{theorem}

See Wakker (1989, Chapter V.5), in particular Theorems V.4.6, V.5.2, and Observation V.3.4'. An even stronger result could be obtained using Wakker (1993).

Kopylov (2010) provides a nice alternative characterization for countable additive subjective expected utility. However, he obtains just non-atomistic probability measures, which does not fit our framework. 

Note that by Theorem~\ref{SEU_theorem}, the Sure Thing Principle must hold: For any event $E \in \Sigma_{\Omega}$ and any $f, g, h, h' \in \mathcal{F}$,
$$f_{E} h \succeq g_{E} h \ \mbox{ if and only if } \  f_{E} h' \succeq g_{E} h'.$$ Thus, we can define for each nonempty $E \in \Sigma_{\Omega}$, a conditional preference relation $\succeq_E$ on $\mathcal{F}$ by for $f, g \in \mathcal{F}$, $$f \succeq_E g \mbox{ if and only if } f_E h \succeq g_E h$$ for some $h \in \mathcal{F}$.

\begin{definition} We say that $\succeq_E$ for nonempty $E \in \Sigma_{\Omega}$ admits a $E$-conditional countable additive subjective expected utility representation when $\succeq$ in Definition~\ref{SEU} is replaced by $\succeq_E$, $\mu$ is replaced by the conditional probability measure $\mu(\cdot \mid E)$, and $u$ is replaced by $u_E$.
\end{definition}

For event $E \in \Sigma_{\Omega}$ and consequences $x, y, z, w \in C$, write $$xy \unrhd_E zw$$ (i.e., subscript $E$) if there exists an event $F \in \Sigma_{\Omega}$ and simple acts $f, g \in \mathcal{F}$ such that $x_F f \succeq_E y_F g$ and $z_F f \preceq_E w_F g$.

\begin{property}[Risk Invariance]\label{risk_invariance} For any events $E, F \in \Sigma_{\Omega}$, $\unrhd_E = \unrhd_F$.
\end{property}

Note that above exposition allows $C = \mathbb{R}$. In such a case, it is handy to also consider a very strict notion monotonicity:

\begin{property}[Strict monotonicity]\label{strict_monotonicty} Let $C = \mathbb{R}$. $\succeq$ is strictly monotone if $x, y \in \mathcal{F}$, $x > y$ if and only if $x \succ y$.
\end{property}

\begin{corollary} Let $C = \mathbb{R}$. For any nonempty $E \in \Sigma_{\Omega}$, $\succeq_E$ Properties~\ref{weak_order} to~\ref{boundedly-strict-continuity} if and only if it admits the $E$-conditional countable additive subjective expected representation. Moreover, $u_E$ can be chosen such that $u_E = u$ if and only if $\succeq_E$ satisfies Property~\ref{risk_invariance}. Finally, $u$ is strictly monotone if and only if $\succeq_E$ satisfies strict monotonicity.
\end{corollary}


\end{document}